\documentclass[pdflatex,sn-mathphys-num]{sn-jnl}

\usepackage{graphicx}%
\usepackage{multirow}%
\usepackage{amsmath,amssymb,amsfonts}%
\usepackage{amsthm}%
\usepackage{mathrsfs}%
\usepackage[title]{appendix}%
\usepackage{xcolor}%
\usepackage{textcomp}%
\usepackage{manyfoot}%
\usepackage{booktabs}%
\usepackage{algorithm}%
\usepackage{algorithmicx}%
\usepackage{algpseudocode}%
\usepackage{listings}%
\usepackage{braket}

\newcommand{\modu}[1]{\left|{#1}\right|}

\newcommand{\trev}{T_{\rm rev}}

\theoremstyle{thmstyleone}%
\theoremstyle{thmstyletwo}%

\theoremstyle{thmstylethree}%

\begin{document}

\title[Tomogram-based quantifiers of nonclassicality dynamics in Kerr and cubic media]{Tomogram-based quantifiers of nonclassicality dynamics in Kerr and cubic media}


\author[1]{\fnm{ K. M. Athira} }

\author[2]{{M. J. Neethu} }

\author*[1,2]{{M. Rohith} }\email{rohith.manayil@gmail.com}

\affil[1]{\orgdiv{P. G. \& Research Department of Physics}, \orgname{Government Arts and Science College Kozhikode},  \orgaddress{\postcode{673018}, University of Calicut, \state{Kerala}, \country{India}}}

\affil[2]{\orgdiv{Quantum Systems Lab, Department of Physics}, \orgname{Government College Malappuram}, \orgaddress{\postcode{676509}, University of Calicut, \state{Kerala}, \country{India}}}


\abstract{The reliable quantification of nonclassicality in quantum states under realistic decoherence remains a central challenge in advancing quantum technologies. Conventional quantifiers such as Wigner negativity, Mandel's $Q$-parameter, nonclassical depth, etc., are often experimentally intractable, non-unique, or insensitive to key quantum signatures. We demonstrate that tomogram-based measures the homodyne nonclassical area and sum tomographic entropy offer a robust, experimentally accessible alternative for quantifying nonclassicality dynamics, as they can be directly obtained from optical tomograms via balanced homodyne detection, avoiding density matrix reconstruction and ensuring feasibility. We study coherent, photon-added coherent, and even coherent states evolving in Kerr and cubic nonlinear systems, with environmental effects modeled using the Lindblad master equation under amplitude and phase damping. The homodyne nonclassical area, which quantifies the excess quadrature variance beyond that of a coherent state, tracks both the onset and decay of nonclassicality, clearly identifying fractional revivals, wave packet splitting, and macroscopic superpositions. We find that amplitude damping drives a rapid monotonic decay toward vacuum, while phase damping allows partial revival features to survive longer. Complementing this, the sum tomographic entropy derived from conjugate-quadrature tomograms captures higher-order fractional revivals and phase-space interference through persistent entropy minima under weak damping. Our results establish homodyne-based quantifiers as powerful, real-time, and experimentally viable tools for tracking nonclassical dynamics in nonlinear optical media, offering a compelling alternative to conventional, experimentally challenging measures.}




\maketitle

\section{Introduction}\label{sec1}

Nonclassicality, defined as the departure of quantum states from classical descriptions, is a fundamental resource for quantum technologies as it enables tasks that are inefficient or impossible with classical states. Nonclassical states facilitate quantum protocols that outperform their classical counterparts. Incorporating nonclassical statistics in quantum operations has enabled breakthroughs in quantum communication, sensing, and computation \citep{Walls1994}. Such states are characterized by nonclassical properties like entanglement \citep{Vogel2014}, squeezing \citep{Walls1983}, sub-Poissonian photon statistics \citep{Mandel1979} and photon antibunching \citep{Kimble1977}. Often the nonclassical properties, namely entanglement and squeezing, are induced by nonlinear phenomena and are widely used for advanced quantum information processing. 

Nonlinearity or nonlinear models are an indispensable resource to generate nonclassicality \citep{Lutkenhaus1995, Albarelli2016}.  The advent of lasers as a strong, coherent source catalyzed in-depth studies into nonlinear optical phenomena. A linear optical process can only manipulate the existing quantum features. Conversely, for the generation of genuinely quantum states from pure classical inputs, nonlinear interactions like the Kerr effect, Parametric down conversion and higher-order wave mixing are required. They impart anharmonicities required to induce distinctly nonclassical characteristics such as entanglement, squeezing and sub-Poissonian statistics \citep{Pezze2018}. Nonlinear effects, including the Kerr effect, triggers an intensity-dependent phase shift to the optical pulse passing through it, making an isotropic medium anisotropic, and leading to  nonclassical phenomena such as revivals and fractional revivals \citep{Agarwal1989, Rohith2014}. While revivals imply the periodic reformation of the state to its initial form, fractional revivals correspond to the splitting of the initial wave packet into multiple, phase-rotated scaled replicas that interfere coherently, forming macroscopic quantum superpositions. Such states are relevant in several quantum information protocols, including quantum cloning \citep{Cerf2000}, and have been extensively investigated in the context of wave-packet dynamics and quantum state transfer \citep{ Xie2023, Chen2007}. Coherent state propagating through a Kerr medium generates Schrodinger cat states, which hold significant potential for application in quantum information and fundamental quantum mechanics \citep{Filho1996, Rohith2015}. Kerr nonlinearity is remarkable in inducing nonreciprocity, allowing directional control over photonic states for optical information processing \citep{Miao2024}. Nonlinear models have become an ideal platform to generate entanglement \citep{Cui2020}, squeezing \citep{Guo2024} and quantum polaritons \citep{Kyriienko2020}. Hence nonlinear processes are widely used for quantum state engineering \citep{Cui2020, Yanagimoto2020, Lamprou2025}.

To effectively harness these advantages, it is essential to identify and quantify nonclassicality, thereby distinguishing genuine quantum states from classical states. This quantification is achieved through nonclassicality quantifiers, which measure the extent of deviation from classicality. In quantum mechanics, the presence of a Gaussian distribution does not guarantee classicality. A squeezed state or an entangled Gaussian state is inherently nonclassical yet still exhibits a Gaussian distribution. Hence one of the important nonclassicality quantifiers was related to the negativity of the phase-space distribution, such as the Wigner Negativity \cite{Wigner1932}. While Wigner negativity is widely used, it is worth noting that some nonclassical states (like squeezed states) possess positive Wigner functions but remain nonclassical. Other phase-space distributions, such as the Glauber-Sudarshan P-function \citep{Glauber1963, Sudarshan1963} or the Husimi Q-function  \citep{Husimi1940}, can provide complementary insights; though the P-function can be highly singular for nonclassical states, the Q-function takes zero values for nonclassical states. Over time, various nonclassicality quantifiers have been introduced to measure the quantumness including Mandel's $Q$-parameter \citep{Mandel1979}, Nonclassical distance \citep{Hillery1987}, nonclassical depth \citep{Lee1991}, Entanglement potential  \citep{Asboth2005}, quantifier based on the number of quantum superpositions \citep{Vogel2014}, negativity volume of the Wigner function \citep{Kenfack2004}, Another measure grounded in the negativity of an observable, which contrasts with its positive semidefinite classical counterpart was also introduced \citep{Gehrke2012},  and Schmidtt number witnesses \citep{Mraz2014}. The experimental certification of nonclassical photonic states has also been an important area of research in recent years, with methods developed for detector-independent verification \citep{Bohmann2019} and advanced state tomography \citep{Gebhart2021}. Nonclassicality has also been experimentally certified for realistic measurement under coarse graining \citep{Roh2023}. Recently, the nonclassical area has been introduced as an alternative quantifier of nonclassicality, defined in terms of the quadrature variance directly obtainable from optical tomograms \cite{Rohith2023}. Optical tomograms, being an alternative representation for quantum states, are positive marginal distribution obtained from the homodyne detection of quadrature phases and hence form a complete quorum \citep{Vogel1989, Lvovsky2009}. This measure is unique for its experimental accessibility, as it bypasses the need for full state reconstruction and can be straightforwardly implemented using optical homodyne tomography. Also, the nonclassical area is robust against decoherence effects, thereby providing reliable verification of nonclassicality dynamics in open quantum systems.

While nonclassicality is a cornerstone of quantum technologies, its robust quantification and dynamical analysis remain challenging, particularly in the presence of nonlinear interactions and environmental decoherence. Traditional quantifiers such as Wigner negativity, Mandel's $Q$-parameter, nonclassical depth, etc., are either experimentally demanding, reconstruction-heavy, or insensitive to certain nonclassical features, limiting their practical applicability. Recent advances, including the use of optical tomograms for state characterization and the introduction of the nonclassical area as a measurable quantity based on quadrature fluctuations, has provided a useful indicator of deviations from coherent-state behavior and offers a practical tool for experimentally probing nonclassical features in specific classes of quantum states. In this work, we use a homodyne-based tomographic framework for real-time tracking of nonclassicality dynamics in nonlinear quantum optical systems, specifically focusing on Kerr and cubic nonlinear media. Specifically, we employ the homodyne nonclassical area and sum tomographic entropy, which can be directly extracted from optical tomograms without density matrix reconstruction. We pay particular attention to revivals and fractional revivals, which naturally emerge in nonlinear media due to anharmonic phase evolution. The tomogram-based quantifiers provide an experimentally accessible and physically intuitive alternative to conventional measures for identifying revivals and fractional revivals in nonlinear media. To evaluate the robustness of nonclassical features under realistic conditions, we model amplitude and phase damping using the Lindblad master equation. This framework enables clear identification of revivals, fractional revivals, and macroscopic quantum superpositions, while also tracking how these features evolve and gradually degrade under decoherence. Thus, our method offers a significant advancement over traditional techniques, combining experimental feasibility with enhanced sensitivity to dynamical nonclassical features. Section~\ref{Sec2} outlines the tomographic framework used to quantify nonclassicality. In section~\ref{Sec3}, we investigate the dynamics of nonclassicality in a Kerr medium, presenting our results on the evolution of the nonclassical area and tomographic entropy under both amplitude and phase damping models. Section~\ref{Sec4} extends the analysis to a cubic nonlinear medium, with a similar exploration of the nonclassicality dynamics. Section~\ref{Sec5} generalizes the important conclusions of the paper.

\section{Tomographic Framework for Quantifying Nonclassicality }\label{Sec2}

To investigate the nonclassicality dynamics in a nonlinear medium, unitary and non-unitary evolution under two distinct nonlinearities incorporating the Kerr Hamiltonian and Cubic Hamiltonian are considered. A Kerr-like medium has an energy spectrum that exhibits a quadratic dependence on the photon number. The effective Hamiltonian of the Kerr-like medium is expressed as:
\begin{equation}
    H_{\rm{kerr}} = \hbar\chi_1 N(N-1) =\hbar\chi_1 {a^{\dagger}}^2 a^2,\label{Hamilt_sys1}
\end{equation}
For a cubic nonlinear medium, the Hamiltonian is given by
\begin{equation}
    H_{\rm{cub}} = \hbar\chi_2 N(N-1)(N-2) =\hbar\chi_2 {a^{\dagger}}^3 a^3.\label{Hamilt_sys2}
\end{equation}
Here, $N$ denotes the number operator, $\chi_1$ and $\chi_2$ correspond to the higher order nonlinear susceptibilities of the medium. The operators $a^\dagger$ and $a$ denote the bosonic creation and annihilation operators for the field, respectively, and satisfy the canonical commutation relation $[a, a^\dagger] = 1$. Throughout this analysis $\hbar$ is set to unity and $\chi_1$ and $\chi_2$ are set to $5$. These Hamiltonians lead to non-equidistant energy level spacing which in turn triggers quantum revivals and nonclassical state generation. 

A general framework to study the non-unitary evolution of open quantum systems due to environment-induced decoherence is given by the Lindblad master equation. The environment is modeled as an ensemble of infinite, non-interacting quantum harmonic oscillators. For a Markovian open quantum system,

\begin{equation}
 \frac{d\rho}{dt} = -i[H, \rho] + \sum_j \left( L_j \rho L_j^\dagger - \frac{1}{2} \left\{ L_j^\dagger L_j, \rho \right\} \right)
\end{equation}
where $\rho(t)$ represents the system's density matrix at time $t$,
\begin{equation}
\rho(t)=\sum_{n,m} \rho_{n,m} \ket{n}\bra{m}.
\label{statematrix}
\end{equation}
Here, $H$ is the system Hamiltonian and $L_j$ are the Lindblad operators representing distinct channels of decoherence. Amplitude damping and phase damping emerge as paradigmatic examples of decoherence through the specific choice of $L_j$.
Under the Born-Markov and secular approximations, the zero-temperature Lindblad master equation is given by
\begin{equation}
    \frac{\partial \rho}{\partial t} = -i\left[H, \rho\right] + \frac{\gamma}{2} \left(2 L\rho L^\dagger - L^\dagger L \rho - \rho L^\dagger L \right).\label{ampdcymaster}
\end{equation}
Amplitude damping model captures energy dissipation to the surroundings such as spontaneous emission or energy relaxation to ground state. This will affect both the population and coherence of the system's density matrix. For the amplitude damping model
\begin{equation}
    \hat{L} = \sqrt{\gamma}\,{a}
\end{equation}
where $\gamma$ is the interaction strength between the system and environment.

In contrast, phase damping involves only the coherence without any energy exchange, hence the population remains intact leading to pure dephasing \citep{Gardiner2004}.
For the phase damping model,
\begin{equation}
    \hat{L} = \sqrt{\gamma} \,N
\end{equation}
The optical tomogram of an arbitrary quantum state  $\rho(t)$ encompasses all the state properties and is expressed as:
\begin{equation}
    \omega(X_{\theta},\theta,t) = \bra{X_{\theta},\theta} \rho(t) \ket{X_{\theta},\theta},\label{optTomo}
\end{equation}
where the state $\ket{X_{\theta},\theta}$ is the eigenstate of the homodyne phase-rotated quadrature operator:
\begin{equation}
    \mathbb{X}_{\theta} = \frac{1}{\sqrt{2}}(a e^{-i\theta} + a^\dagger e^{i\theta}),
\end{equation}
with eigenvalue $X_{\theta}$. Here, $\theta \in [0, 2\pi]$ represents the phase of the local oscillator in the homodyne detection setup. The eigenstate $\ket{X_{\theta},\theta}$ in terms of the ground state $\ket{0}$ is given by:
\begin{equation}
\ket{X_\theta,\theta} = \frac{1}{\pi^{1/4}} \exp\left[-\frac{{X_\theta}^2}{2} - \frac{1}{2}e^{i2\theta}{a^\dag}^2 + \sqrt{2}e^{i\theta}{X_\theta}{a^\dag}\right]\ket{0}
\end{equation}
For a pure quantum state $\ket{\psi}$, the optical tomogram becomes:
\begin{equation}
    \omega(X_{\theta},\theta) = |\bra{X_{\theta},\theta}{\psi}\rangle|^2.
\end{equation}
The optical tomogram satisfies the normalization condition:
\begin{equation}
    \int_{-\infty}^{\infty} dX_\theta \, \omega(X_{\theta},\theta) = 1,
\end{equation}
and exhibits the reflection property:
\begin{equation}
    \omega(X_{\theta},\theta + \pi) = \omega(-X_{\theta},\theta).
\end{equation}
Furthermore, the $n$-th order moments of the quadrature operator can be calculated directly from the optical tomogram using the following relation:
\begin{equation}
\langle \mathbb{X}_{\theta}^{n}  \rangle = \int_{-\infty}^{\infty} dX_{\theta} \, X_{\theta}^n \omega(X_{\theta},\theta)
\label{moment}    
\end{equation}
The nonclassical area serves as a sensitive indicator of deviations in the optical tomogram by quantifying the difference in tomographic area relative to that of a reference coherent state, thereby capturing non-Gaussian and interference-induced structural features in the quadrature distribution. If $\sigma(\ket{\psi})$ represents the area projected by an optical tomogram onto the $X_{\theta}-\theta$, then nonclassical area of a state is defined as:
\begin{equation}
\sigma(\ket{\psi}) - \sqrt{2}\pi = \int_{0}^{2\pi} \Delta X_{\theta} \, d\theta - \sqrt{2}\pi,
\label{nca}    
\end{equation}
where $\Delta X_{\theta} = \sqrt{\langle \mathbb{X}_{\theta}^2 \rangle - \langle \mathbb{X}_{\theta} \rangle^2}$ is the standard deviation of the rotated quadrature operator $\mathbb{X}_{\theta}$. 

For pure single-mode states, a nonzero value of the nonclassical area serves as a sufficient indicator of deviations from coherent-state behavior and may signal the presence of nonclassical features. However, this quantity does not constitute a strict nonclassicality criterion in general, as classical mixed states (e.g., thermal state) may also yield nonzero values due to statistical mixing. The magnitude of this quantity generally increases with the strength of nonclassicality-inducing operations such as squeezing or photon addition \citep{Rohith2023}.

The nonclassical area of a quantum state is closely related to the information content associated with the quadrature operator, specifically through the concept of Wigner-Yanase skew information \cite{Wigner1997}. For a pure state $\rho = \ket{\psi}\bra{\psi}$, the variance of the quadrature operator simplifies to the skew information of a bosonic field state.

The optical tomogram, being a positive probability distribution, has entropy and information pertained with it. A time $t$ the tomographic entropy associated with an optical tomogram $\omega(X_{\theta}, \theta, t)$, along quadrature angle 
\(\theta\) is given by:
\begin{equation}
    S(\theta) = -\int dX_{\theta} \, \omega(X_\theta, \theta,t) \ln \omega(X_{\theta}, \theta, t) \label{tomogramentropy}
\end{equation}
It satisfies the uncertainty relation:
\begin{equation}
    S(\theta) + S(\theta + \pi/2) \geq 1 + \ln \pi.
    \label{entropy_uncertainty}
\end{equation}
For pure minimum-uncertainty Gaussian states, the entropic uncertainty relation attains equality for appropriate choices of quadrature angles.
In our analysis, we utilize both the nonclassical area as a measure of nonclassicality and the tomographic entropy of the state to investigate the time evolution of a wave packet in a Kerr and cubic nonlinear medium. 

The system we consider has infinite-dimensional Hilbert space. But in real computation we have to choose a Hilbert space dimension which is sufficient enough to ensure convergence. The required Hilbert space dimension for calculating the nonclassical area and tomographic entropy was determined through various convergence checks and trial and error method. Specifically, convergence was tested by increasing the cutoff dimension $n_{\rm{max}}$ and monitoring the stability of relevant physical observables upon increasing $n_{\rm{max}}$. It was confirmed that these quantities remain unchanged within a tolerance of $10^{-10}$ beyond a certain cutoff. In addition, the trace of the density matrix is preserved throughout the evolution with deviations less than $10^{-10}$, and the population near the truncation boundary remains negligible at all times.

\section{Nonclassicality Dynamics in a Kerr medium}\label{Sec3}

In this section, we present our investigation of the nonclassicality dynamics using the homodyne nonclassical area and tomographic entropy. We first consider a Kerr-like medium as the nonlinear model through which the wave packet propagates. The analysis is initially carried out for a closed system ($\gamma = 0$). In the subsequent subsections, we extend the study to open-system dynamics by examining the nonclassical area and tomographic entropy under amplitude and phase damping models of decoherence, by invoking a weak coupling between the system and its environment ($\gamma \ll 1$). 

The Kerr-like medium, described by the Hamiltonian in Eq.~(\ref{Hamilt_sys1}), exhibits full quantum revivals as a result of quantum interference effects, which occur when the evolution time is an integer multiple of the revival time, $T_{rev} = \pi/\chi_1$. During the evolution period $0 \leq t \leq T_{rev}$, the wave packet at specific times evolves into a superposition of $l$ phase-rotated and scaled replicas of the original wave packet, referred to as $l$-subpacket fractional revivals. The timing of these fractional revivals is influenced not only by the system's Hamiltonian but also by the characteristics of the initial wave packet. For an initial coherent state $\ket{\alpha}$, the $l$-subpacket fractional revivals occur at times $t = j T_{rev}/l$, where $j = 1, 2, \dots, (l-1)$ within the interval $0 \leq t \leq T_{rev}$. Coherent states, with a Fock state expansion of the form $\sum_n C_n \ket{n}$, exhibit this behaviour, however, for initial wave packets with a Fock state expansion of the form $\sum_n D_{2n} \ket{2n}$, such as even coherent states, the $l$-subpacket fractional revival times occur at $t = j T_{rev}/4l$, where $j = 1, 2, \dots, (4l-1)$ \citep{Rohith2015}.

 We study the nonclassicality dynamics with the tomogram-based metrics, the nonclassical area and tomographic entropy, for three classes of initial states: a coherent state, a 3-photon added coherent state and an even coherent state evolving within a Kerr medium. Coherent states, defined as the eigenstates of the annihilation operator, satisfy the relation \( a \ket{\alpha} = \alpha \ket{\alpha} \), where \(\alpha\) is a complex parameter expressed as \(\alpha = |\alpha| e^{i\delta}\). 
\begin{equation}
C_n=e^{-\modu{\alpha}^2/2}\frac{\alpha^n}{\sqrt{n!}} 
\end{equation} 
These states can closely approximate the quantum state of a well-tuned laser. The optical tomogram associated with such a state is given by \cite{Rohith2015}:
\begin{equation}
    \omega_{\alpha} \left(X_\theta, \theta\right) = \frac{1}{\sqrt{\pi}} \exp\left\{-\left(X_\theta - \sqrt{2} |\alpha| \cos(\delta - \theta)\right)^2\right\}.
\end{equation}
In the optical tomographic plane (i.e., the \(X_\theta - \theta\) plane), this tomogram exhibits the structure of a single sinusoidal strand. We then consider the evolution of an initial \(p\)-photon-added coherent state, defined as
\begin{equation}
\ket{\alpha, p} = N_{\alpha,p}\left(a^\dagger\right)^p \ket{\alpha},
\end{equation}
where \(p\) is a positive integer, \(N_{\alpha,p}\) is the normalization constant, and \(\ket{\alpha}\) denotes a coherent state. 
For \(p \leq n\)
\begin{equation}
C_n= \frac{e^{-\modu(\alpha)^2/2} \alpha^{n-p}\sqrt{n!}}{ \sqrt{p! L_p(-\modu{\alpha}^2) (n-p)!}}
\end{equation}
where \(L_p(-|\alpha|^2)\) is the Laguerre polynomial of order p.
We also investigate the evolution of  an initial even coherent state, defined as the superposition of two coherent states with a \(\pi\)-phase difference:  
\begin{equation}
\ket{\alpha}_+ = N_+ \left[ \ket{\alpha} + \ket{-\alpha} \right],
\end{equation}
where \(N_+\) is the normalization constant. The Fock state expansion coefficient for an even coherent state is
\begin{equation}
C_n = 2 N_{+} \, e^{-|\alpha|^2/2} \, \frac{\alpha^n}{\sqrt{n!}} \, 
\delta_{\left[\frac{n}{2}\right], \frac{n}{2}}
\end{equation}
where \([.]\) denotes the integer part and \(\delta\) is the kronecker-delta function. The tomogram-based nonclassicality analysis of the system can be extended to other forms of initial wave packets. We use the optical tomograms of the above mentioned states to calculate the homodyne nonclassical area (using Eq. (\ref{moment}) \& Eq. (\ref{nca})) and tomographic entropy (using Eq. (\ref{tomogramentropy})) in a Kerr-like nonlinear medium. 

\subsection{Dynamics of nonclassical area}\label{Sec3.1}

\begin{figure}[h]
\centering
\includegraphics[width=0.5\textwidth]{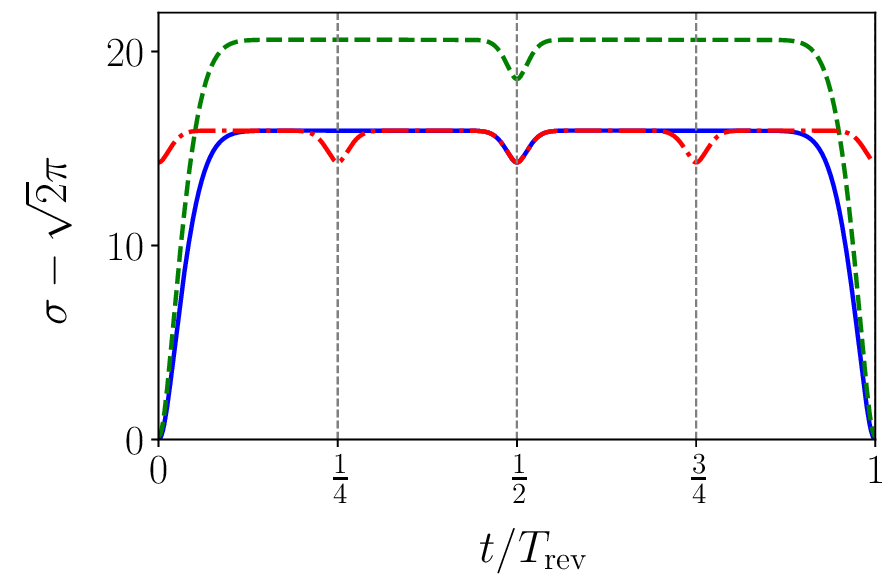}
\caption{Nonclassical area as a function of time for an initial coherent state (solid line), a $3$-photon-added coherent state (dashed line), and an even coherent state (dash-dot line) for a field strength $\modu{\alpha}^2=10$ and $\gamma=0$ with a Hilbert space truncation of $n_{\rm{max}}=60$.}
\label{fig1}
\end{figure}

We proceed to investigate the evolution of the nonclassical area associated with the time-dependent state \(\rho(t)\), obtained from the solution of the master equation Eq.~(\ref{ampdcymaster}), utilizing Eq.~(\ref{optTomo}) for a range of initial states of the system. The nonclassical area quantifies the effective region mapped by the optical tomogram of the evolving state onto the optical tomographic plane, surpassing the extent corresponding to a coherent state. For a unitary evolution in the Kerr medium, the state at instant $t$ is,
\begin{equation}
\rho_{nm}(t)= e^{-i\chi(n(n-1)-m(m-1))t}\rho_{nm}(0)
\end{equation}
The overlap between the quadrature eigenstate and the Fock state is given by,
\begin{equation}
    \langle X_\theta, \theta | n \rangle = \frac{1}{\pi^{1/4}}\frac{1}{2^{n/2} \sqrt{n!}} \, e^{-X_\theta^2/2} H_n(X_\theta) e^{-in\theta}
    \label{fock_quad}
\end{equation}
where $H_n(X_\theta)$ is the Hermite polynomial of $n^{th}$
order. Ergo, the optical tomogram for the state under unitary evolution can be expressed using Eq.~(\ref{statematrix}), Eq.~(\ref{optTomo}) \& Eq.~(\ref{fock_quad}) as,
\begin{eqnarray}
\omega(X_\theta, \theta, t) & =\frac{e^{-X_{\theta}^2}}{\sqrt{\pi}}\sum_{n,m=0}^{\infty}
\frac{H_n(X_{\theta})H_m(X_{\theta})}{2^{(n+m)/2}\sqrt{n!m!}}\nonumber\\
& e^{-i(n-m)\theta}\,e^{-i\chi(n(n-1)-m(m-1))t}
\rho_{n, m}(0).
\end{eqnarray}
The optical tomogram can be used to compute the nonclassical area at any instant $t$. 
Figure~\ref{fig1} depicts the dynamics of nonclassical area in the absence of decoherence (\(\gamma=0\)) for three different initial states: coherent state, 3-photon added coherent state and an even coherent state evolving through a Kerr medium. We have verified that, for all three initial states, $\rho^2 = \rho$ at $t = 0$, confirming the purity of the states. During the evolution, however, the state becomes mixed, such that $\mathrm{Tr}(\rho^2) < 1$.
In figure~\ref{fig1}, the variation of the nonclassical area with time is shown for \(t \leq \trev\), with the dynamics repeating precisely after this interval.  When no decoherence is involved, the evolution is simply as per the Schrodinger equation, showing periodic revivals. Though a coherent state \(\ket{\alpha}\) has its initial nonclassical area zero, Kerr nonlinearity induces nonclassicality to the state and is reflected in the corresponding nonclassical area. For the initial $3$-photon-added coherent and even coherent states, the nonclassical area starts from a nonzero value, reflecting their inherently nonclassical nature. 

The nonzero initial values of the nonclassical area reappear at the revival time \(\trev\).  These results reveal a periodic behaviour, characterized by a recurrence at \(t = \trev\), where the nonclassical area returns to its initial value at \(t = 0\), irrespective of the initial state. This periodic recurrence serves as a definitive indicator of wave packet revivals. Local minima observed periodically in the nonclassical area dynamics correspond to macroscopic superpositions of states. The number of minima indicates the number of times the two-subpacket fractional revivals occur. This behavior fundamentally arises because the nonclassical area is based on the second moment of the rotated quadrature operator. These local minima signify a reduced phase-space spread, which is a hallmark of nonclassicality.

\begin{figure}[h]
\centering
\includegraphics[width=0.45\textwidth]{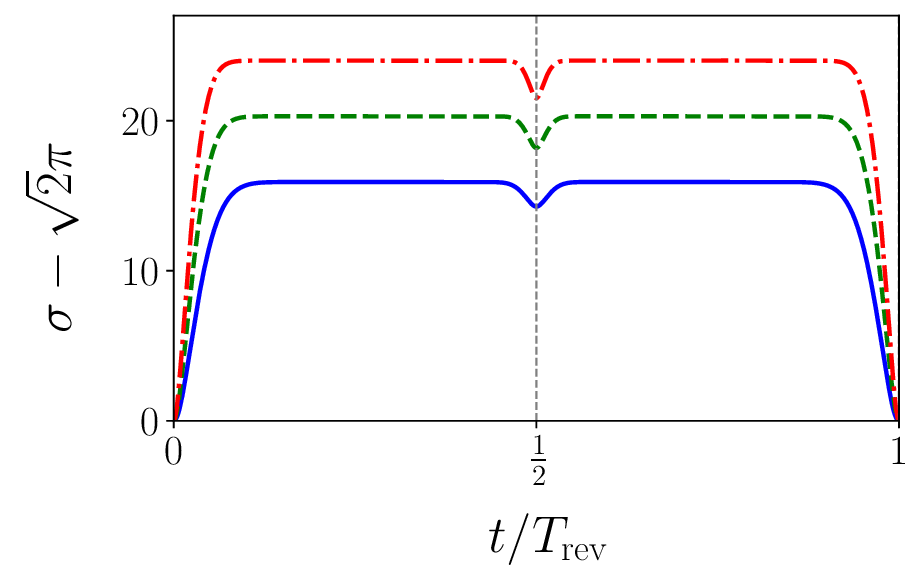}
\caption{Time evolution of the nonclassical area in the absence of decoherence \(\gamma = 0\) for an initial coherent state with mean photon numbers $\modu{\alpha}^2=10$   (solid line), $\modu{\alpha}^2=15$ (dashed line), and $\modu{\alpha}^2=20$ (dash-dot line).}
\label{fig2}
\end{figure}

Between \(t=0\) and \(t=\trev\), the nonclassical area exhibits a local minimum at \(t=\trev/2\) for all initial states, signifying a two-subpacket fractional revival for \(\ket{\alpha}\) and \(\ket{\alpha,3}\) as the wave packet evolves through the Kerr-like medium. In contrast, for the even coherent state \(\ket{\alpha}_+\), the minima at \(t=\trev/4\), \(t=\trev/2\), and \(t=3\trev/4\) correspond to phase space rotations.   Additionally, it is observed that as the field strength \(\modu{\alpha}^2\) increases, the value of the nonclassical area at any given time between $t=0$ and $t=\trev$ increases for all the initial states taken (see figure~\ref{fig2} for the case of initial coherent state). 

Next, we investigate the impact of decoherence on the dynamics of the nonclassical area for various initial states. There are two types of decoherence models involved- the Amplitude damping model and the phase damping model. 

\subsection{Amplitude Decay Model}\label{Sec3.2}
For zero temperature conditions, the amplitude decay model follows an evolution as per the Lindblad master equation given in Eq.~(\ref{ampdcymaster}).
When a state evolves in a Kerr medium subjected to amplitude damping, the state at time $t$ can be expressed using the Kraus operator representation as
\begin{equation}
\rho(t)=\sum_{k=0}^{\infty}\frac{(1-e^{-\gamma t})^k}{k!}L_t U(t)a^k \rho(0){a^\dagger}^k U^\dagger (t)L_t^\dagger
\end{equation}
where $L_t=e^{-\gamma t a^\dagger a/2}$ and $U(t)=e^{-i\chi tn(n-1)}$\\
$\frac{(1-e^{-\gamma t})^k}{k!}$ is the probability for losing $k$ photons. Then the time derivative of the system can be written as
\begin{eqnarray}
\dot{\rho}_{n,m}(t)&=&\left(-i\chi (n(n-1)-m(m-1))\right.\nonumber\\&&
\left.-\frac{\gamma}{2} (n+m)\right)\rho_{n,m}(t) \nonumber\\&&
+\gamma \sqrt{(n+1)(m+1)}\rho_{n+1,m+1}(t).
\end{eqnarray}
Density matrix elements at time t will be
\begin{eqnarray}
\rho_{n,m}(t) &=& 
e^{-i\chi [n(n-1) - m(m-1)]t} \,
e^{-\gamma (n + m)t /2} \nonumber \\
&&\times \sum_{k=0}^{\infty} 
\binom{n + k}{k}^{1/2} \binom{m + k}{k}^{1/2}\nonumber \\ &&\frac{(1 - e^{-\gamma t})^k}{k!} 
\rho_{n + k, m + k}(0).
\end{eqnarray}
Thus, the optical tomogram at instant $t$ has the form,
\begin{eqnarray}
\omega(X_\theta, \theta, t) & =\frac{e^{-X_{\theta}^2}}{\sqrt{\pi}}\sum_{n,m=0}^{\infty}
\frac{H_n(X_{\theta})H_m(X_{\theta})}{2^{(n+m)/2}\sqrt{n!m!}}\nonumber\\
& e^{-i(n-m)\theta}\,e^{-i\chi(n(n-1)-m(m-1))t}\nonumber\\
& e^{-\gamma(n+m)t/2}\sum_{k=0}^{\infty} 
\binom{n + k}{k}^{1/2} \binom{m + k}{k}^{1/2} \nonumber \\ 
& \frac{(1 - e^{-\gamma t})^k}{k!} 
\rho_{n + k, m + k}(0).
\label{optomo_kerr_ampdcy}
\end{eqnarray}
Using the above optical tomogram, nonclassical area dynamics for different states can be studied.
\begin{figure}[h]
\centering
\includegraphics[width=0.5\textwidth]{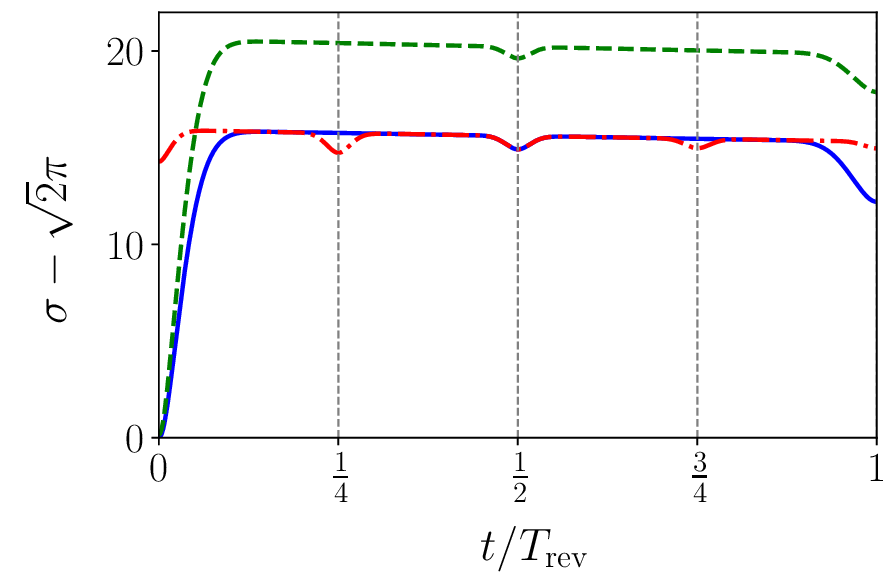}
\caption{Time evolution of the nonclassical area with amplitude damping for \(\gamma = 0.1\) and \(\modu{\alpha}^2=10\), shown for an initial coherent state (solid line), a 3-photon-added coherent state (dashed line), and an even coherent state (dash-dot line). Hilbert space dimension is set to 60.}
\label{fig3}
\end{figure}
Figure~\ref{fig3} illustrates the nonclassical area as a function of time under amplitude damping for a coherent state, 3-photon added coherent state and even coherent state with a coupling strength of \(\gamma = 0.1\). Amplitude damping works as a photon loss channel, showing no exact revivals in the nonclassical area dynamics. Though dips are still observable under weak coupling, the initial value of the nonclassical area is not retained as the system evolves under amplitude decay.
It is obvious from the Eq.~(\ref{optomo_kerr_ampdcy}) that in the asymptotic limit $\gamma t \to \infty$ the system approaches vacuum state, regardless of its initial state since no thermal photons were present initially in the zero temperature case. The corresponding optical tomogram is expressed as  
\begin{equation}  
\omega \left(X_\theta, \theta\right) = \frac{1}{\sqrt{\pi}} e^{-X_\theta^2}.  
\end{equation} 
\begin{figure}[h]
\centering
\includegraphics[width=0.5\textwidth]{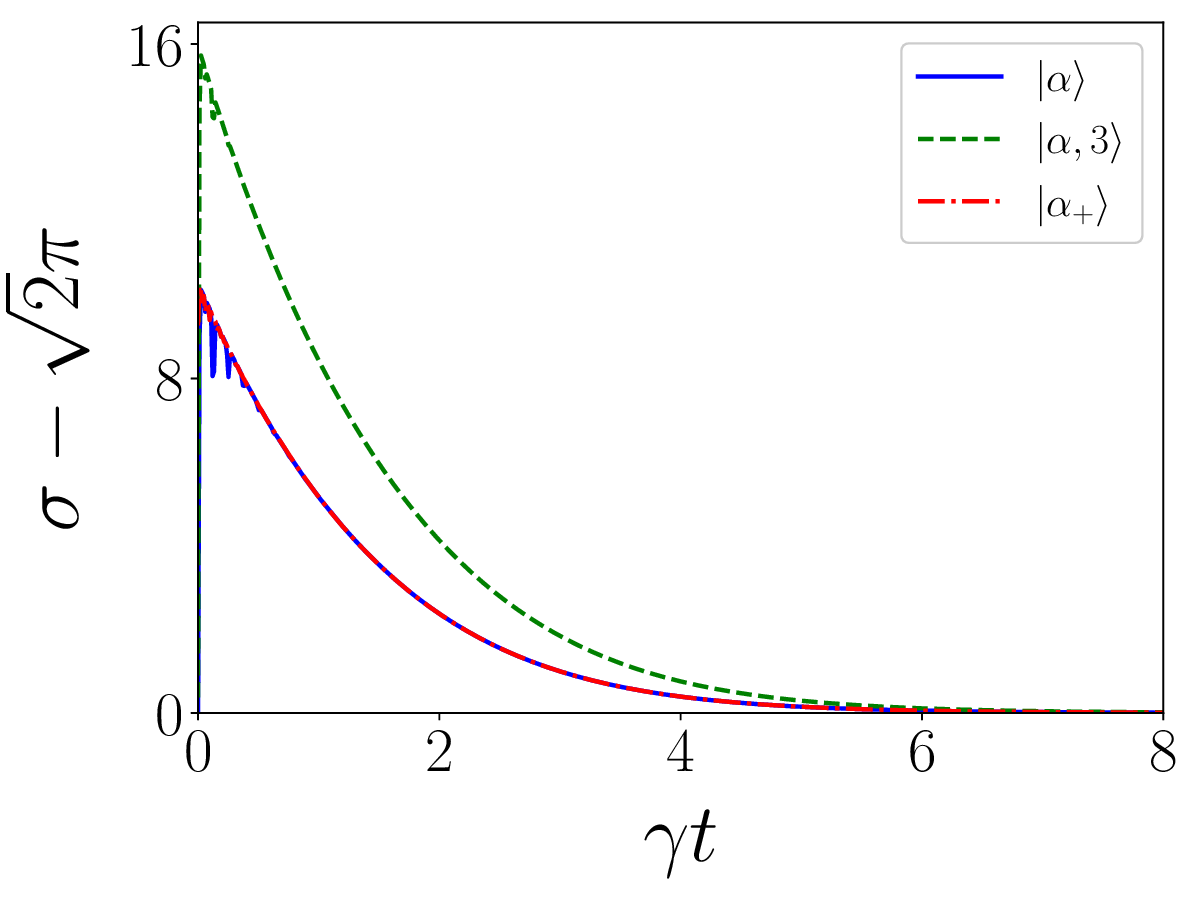}
\caption{Evolution of the nonclassical area as a function of \(\gamma t\) with amplitude damping shown for an initial coherent state (solid line), a 3-photon-added coherent state (dashed line), and an even coherent state (dash-dot line) with \(\modu{\alpha}^2=5\)}
\label{fig4}
\end{figure}

This can be observed in Figure~\ref{fig4} as the nonclassical area exponentially decays to zero, losing both coherence and population. 

\subsection{Phase Decay Model}\label{Sec3.3}
The evolution of a state at zero temperature under phase decay is governed by the master equation  [Eq.~(\ref{ampdcymaster})]. Then the time-evolved state at $t$ is given by
\begin{equation}
\rho_{nm}(t)= e^{-i\chi(n(n-1)-m(m-1))t}e^{-\gamma(n-m)^2 t/2}\rho_{nm}(0).
\end{equation}
When $n=m$, $\rho_{nn}(t)=\rho_{nn}(0)$. Thus the phase decay will only destroy the off-diagonal terms. 
The optical tomogram for a state evolving in a Kerr medium under phase damping is,
\begin{eqnarray}
\omega(X_\theta, \theta, t) & =\frac{e^{-X_{\theta}^2}}{\sqrt{\pi}}\sum_{n,m=0}^{\infty}
\frac{H_n(X_{\theta})H_m(X_{\theta})}{2^{(n+m)/2}\sqrt{n!m!}}\nonumber\\
& e^{-i(n-m)\theta}\,e^{-i\chi(n(n-1)-m(m-1))t}\nonumber\\
& e^{-\gamma(n-m)^2 t/2}\rho_{nm}(0).
\end{eqnarray}
\begin{figure}[h]
\centering
\includegraphics[width=0.5\textwidth]{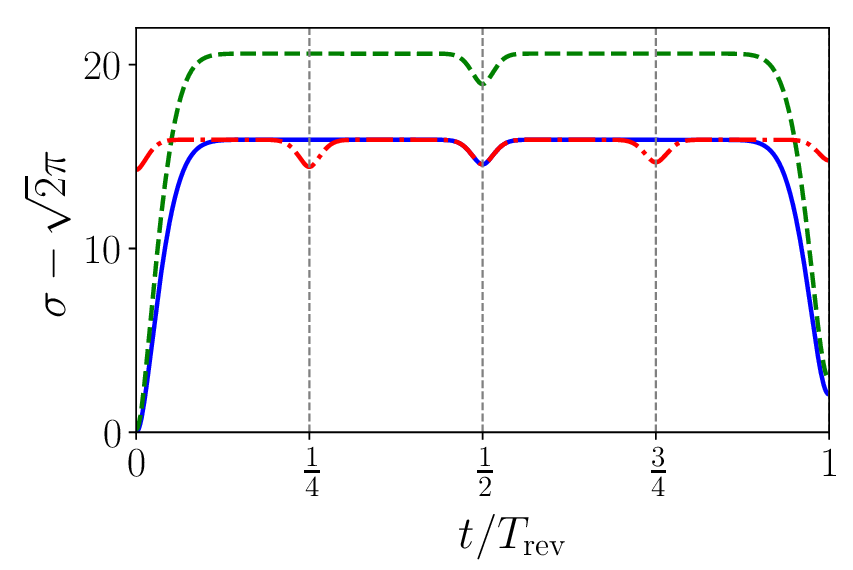}
\caption{Time evolution of the nonclassical area with phase damping for \(\gamma = 0.1\) and \(\modu{\alpha}^2=10\), shown for an initial coherent state (solid line), a 3-photon-added coherent state (dashed line), and an even coherent state (dash-dot line). Fock cutoff is set to 60.}
\label{fig5}
\end{figure}
\begin{figure}[h]
\centering
\includegraphics[width=0.5\textwidth]{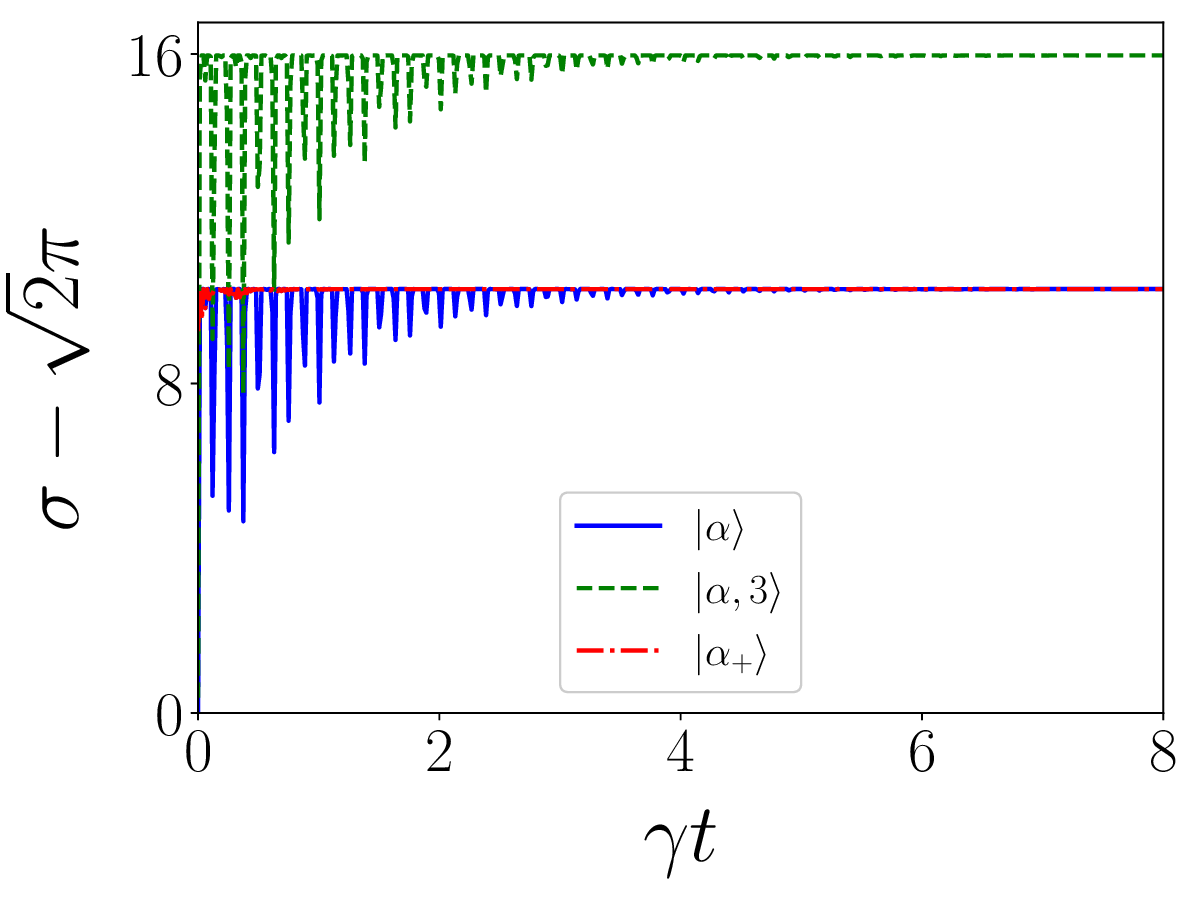}
\caption{Evolution of the nonclassical area as a function of \(\gamma t\) with phase damping shown for an initial coherent state (solid line), a 3-photon-added coherent state (dashed line), and an even coherent state (dash-dot line) with \(\modu{\alpha}^2=5\)}
\label{fig6}
\end{figure}
Figure~\ref{fig5} shows the nonclassical area dynamics under pure dephasing with a coupling strength of \(\gamma = 0.1\).
 Our analysis reveals that, in the presence of decoherence, the nonclassical area does not return to its initial value at time $t=\trev$, signifying the absence of exact revivals during the dissipative evolution of a wave packet in a Kerr-like nonlinear medium. However, even under weak environmental coupling, signatures of near fractional revivals and phase-space rotation persist, evidenced by a local minimum in the nonclassical area at \(t = \trev/2\). This robustness highlights the enduring nature of these features despite environment-induced decoherence. As amplitude damping affects both coherence and population of the system, it contributes to a larger decay compared to phase damping, as observed from figure~\ref{fig3} and figure~\ref{fig5}. Over longer time scales, as $\gamma t \rightarrow \infty$, revivals and two-subpacket fractional revivals gradually fade due to coherence loss, and the nonclassical area eventually stabilizes, as shown in
figure~\ref{fig6}. 

The asymptotic state is given by
\begin{equation}
\rho(\infty) = \sum_n \rho_{nn}(0)\, |n\rangle \langle n|,
\end{equation}
which is strictly diagonal in the Fock basis.
For an initial coherent state,
\[
\rho_{nn}(0)
=  e^{-|\alpha|^2} \frac{|\alpha|^{2n}}{n!} 
\]
In this context, it is important to note that the nonclassical area cannot be regarded as a strict witness of nonclassicality. In the long-time limit under phase damping, the state becomes a phase-randomized coherent-state mixture, due to classical mixing, and the standard deviation of the rotated quadrature operator contains contributions from both classical and quantum uncertainties. Consequently, the nonclassical area remains positive even in this limit~\cite{Rohith2023}, and therefore cannot be treated as a strict nonclassicality witness.
The steady state optical tomogram under phase damping can be then obtained as, 
\begin{equation}  
\omega \left(X_\theta, \theta\right) = \frac{e^{-X_\theta^2}}{\sqrt{\pi}}  \sum_{n=0}^{\infty}\frac{H_n^2(X_\theta)}{2^n n!} \rho_{n,n}(0).  
\end{equation} 
It can also be observed from figure~\ref{fig4} and figure~\ref{fig6} that among the three different states under consideration, the photon-added coherent state is more resilient to damping. While the nonclassical area dynamics effectively capture wave packet revivals and two-subpacket fractional revival signatures, detecting higher-order fractional revivals requires alternative measures, such as analyzing tomographic entropy dynamics.

\subsection{Dynamics of tomographic entropy}\label{Sec3.4}

Nonclassical area dynamics have been shown to reveal two-subpacket fractional revivals and phase-space rotations. For higher-order fractional revivals, which manifest as intricate multi-stranded structures in tomograms, the sum of information entropies associated with the wave packet's probability densities provides a viable alternative, as it encapsulates the informational content of each mini subpacket, irrespective of spatial distribution. 
We employ the tomographic entropy to characterize fractional revivals. Tomographic entropy, a direct measure of the information content, provides insights into the localization and delocalization of the wave packet in phase space.

A minimal entropy in position space indicates a highly confined state, while higher entropy corresponds to a more dispersed state; the same principle applies in momentum space, reflecting momentum uncertainty. Consequently, the sum of tomographic entropies in conjugate spaces (denoted as $\theta$ and $\theta+\pi/2$) serves as a robust metric for quantifying the state's localization and delocalization, thereby complementing the information contained in observed tomograms.
For an initial coherent state \(\ket{\alpha}\) propagating through a Kerr medium without decoherence, the tomogram at instant t is
\begin{eqnarray}
\omega_\alpha(X_\theta, \theta, t) &=& \frac{e^{-|\alpha|^2} \, e^{-X_\theta^2}}{\sqrt{\pi}}\times  \nonumber \\
&& \left| \sum_{n=0}^{\infty} \frac{\alpha^n \, e^{-i \chi t n(n-1)}}{n! \, 2^{n/2}} 
\, e^{-i n \theta} \, H_n(X_\theta) \right|^2.
\end{eqnarray}
For an initial $p$-photon added coherent state 
\(\ket{\alpha,p}\) in a Kerr medium,

\begin{eqnarray}
\omega_{\alpha,p}&&(X_\theta, \theta, t) = 
\frac{e^{-|\alpha|^2}}{p! \, L_p(-|\alpha|^2)} \cdot \frac{e^{-X_\theta^2}}{\sqrt{\pi}} \nonumber \\
&&\times \bigg|\sum_{n=p}^{\infty} 
\frac{\alpha^{n-p} \, e^{-i \chi t n(n-1)}}{(n-p)! \, 2^{n/2}} 
 e^{-i n \theta}  H_n(X_\theta)\bigg|^2.
\end{eqnarray}
An even coherent state \(\ket{\alpha_+}\) evolving inside a Kerr medium has optical tomogram,

\begin{eqnarray}
\omega_{+}(X_\theta, \theta, t) &=& 
4 N_{+}^2 \, \frac{e^{-|\alpha|^2} \, e^{-X_\theta^2}}{\sqrt{\pi}} \times \bigg| \sum_{n=0}^{\infty} 
\frac{\alpha^n \, e^{-i \chi t n(n-1)}}{n! \, 2^{n/2}}  \nonumber \\
&&
\times e^{-i n \theta} \, H_n(X_\theta) \, \delta_{\left[\frac{n}{2}\right], \frac{n}{2}} 
\bigg|^2.
\end{eqnarray}
\begin{figure*}[h]
    \centering
    \includegraphics[width=0.8\textwidth]{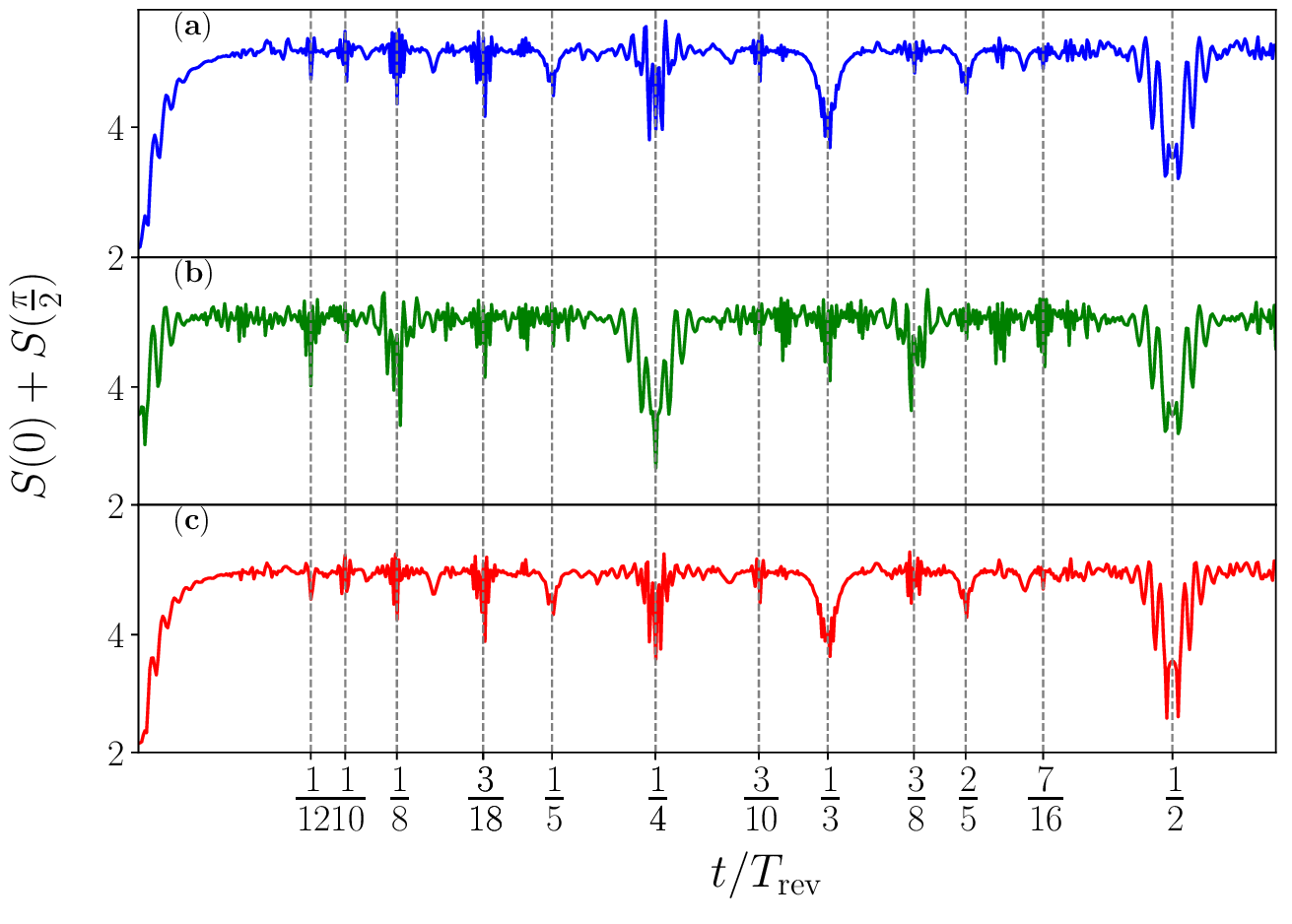}    
    \caption{Dynamics of the sum of tomographic entropies in conjugate spaces for an initial (a) coherent state, (b) three-photon-added coherent state, and (c) even coherent state with $|\alpha|^2=40$ and $n_{\rm{max}}=100$, in the absence of decoherence ($\gamma=0$). Relative minima in the curves correspond to the instants of fractional revivals and phase-space rotations. }
    \label{fig7}
\end{figure*}
In the absence of decoherence, the sum of tomographic entropy in conjugate spaces returns to its initial value at the instants of wave packet revivals, reflecting the retrieval of the initial state. At the instants of fractional revivals, where macroscopic superposition states are formed, the sum of tomographic entropy exhibits relative minima in its temporal dynamics. Here, we analyze the dynamics of the sum of tomographic entropies corresponding to the position space [$\theta=0$ in Eq.~(\ref{tomogramentropy})] and momentum space [$\theta=\pi/2$ in Eq.~(\ref{tomogramentropy})]. Alternatively, the entropy sum can be evaluated for any conjugate spaces $\theta$ and $\theta+\pi/2$, provided both angles lie within $[0,\,2\pi]$. For an initial coherent state, the entropies in the conjugate spaces satisfy $S(\theta) = S(\theta+\pi/2) = \left(1 + \ln\,\pi\right)/2$, leading to a sum $S(\theta) + S(\theta+\pi/2) = 1 + \ln\,\pi$, consistent with minimum uncertainty. In the absence of decoherence ($\gamma=0$), this value is retained at times $t = n\trev$, where $n = 1, 2, 3, \dots$. 
All numerical calculations in this section were performed in a truncated Fock basis. For the three different initial states with amplitude $|\alpha|^2 = 40$ the photon number distribution is checked to be well contained within the Hilbert space dimension $n_{\rm{max}}=100$. It is also verified that the entropy of the coherent state matches the theoretical value $1+\rm{ln}\, \pi$ for the given $n_{\rm max}$ with an accuracy of one part in $10^{6}$. The optical tomogram $\omega(X_{\theta} ,\theta)$ was evaluated over the quadrature range $X_{\theta} \in [-10, 10]$ using a uniform grid of 200 points. The time evolution was computed for the interval $0 \leq t \leq 0.55 \trev$ with 700 uniformly spaced time steps. The entropy sum $S(0) + S(\pi/2)$ was obtained from the corresponding quadrature distributions for coherent, even coherent, and three- photon-added coherent states.

 Figure~\ref{fig7} presents the temporal evolution of the sum of tomographic entropies in conjugate spaces as a function of $t/\trev$ for three initial states: a coherent state, a three-photon-added coherent state, and an even coherent state. The plot is shown up to $t/\trev = 1/2$ because the dynamics are symmetric about $t/\trev = 1/2$. The figure reveals relative minima in the entropy dynamics, indicated by vertical dotted lines, which correspond to the instants of fractional revivals and phase-space rotations. While only the most prominent minima are marked, additional relative minima associated with fractional revivals are discernible in the plot. Furthermore, the signatures of fractional revivals become more pronounced at higher field strengths, enabling the observation of higher-order fractional revivals.

The entropy dynamics successfully capture even higher-order fractional revivals, such as the 18th-order fractional revival. For an initial coherent state, the state at time $3\trev/18$ is a superposition of 18 phase-rotated coherent states. Nonclassicality-inducing operations, such as photon addition or the creation of superposition states, render the state non-Gaussian. Consequently, photon-added coherent states and even coherent states exhibit higher values of the entropy sum in conjugate spaces compared to a coherent state. As these states evolve through the Kerr-like medium, their entropy sum-and hence their nonclassicality-further increases. As noted in earlier sections, the time-evolved state at $t = \trev/4$ for an initial even coherent state is a phase-rotated version of the initial state. This is observed as a relative minimum in the entropy dynamics. The fractional revivals of initial even coherent states at $t = \trev/8$, $\trev/12$, and $7\trev/16$, corresponding to two-, three-, and four-subpacket fractional revivals, respectively, are marked in the entropy dynamics plot. This demonstrates that the dynamics of tomographic entropy serve as a powerful tool for identifying higher-order fractional revivals. The depth of relative minima in the entropy plot depends on the overlap of fractional revivals and phase-space rotations at a given time. For instance, at $t = \trev/4$, the observed minimum corresponds not only to a phase-space rotation of the initial even coherent state but also to fractional revivals such as $3\trev/12$, $4\trev/16$, $5\trev/20$, and others. To precisely determine the specific fractional revival occurring at a given instant, one must analyze the corresponding optical tomogram and apply the strand-counting method.
\begin{figure*}[h]
    \centering
    \includegraphics[width=0.8\textwidth]{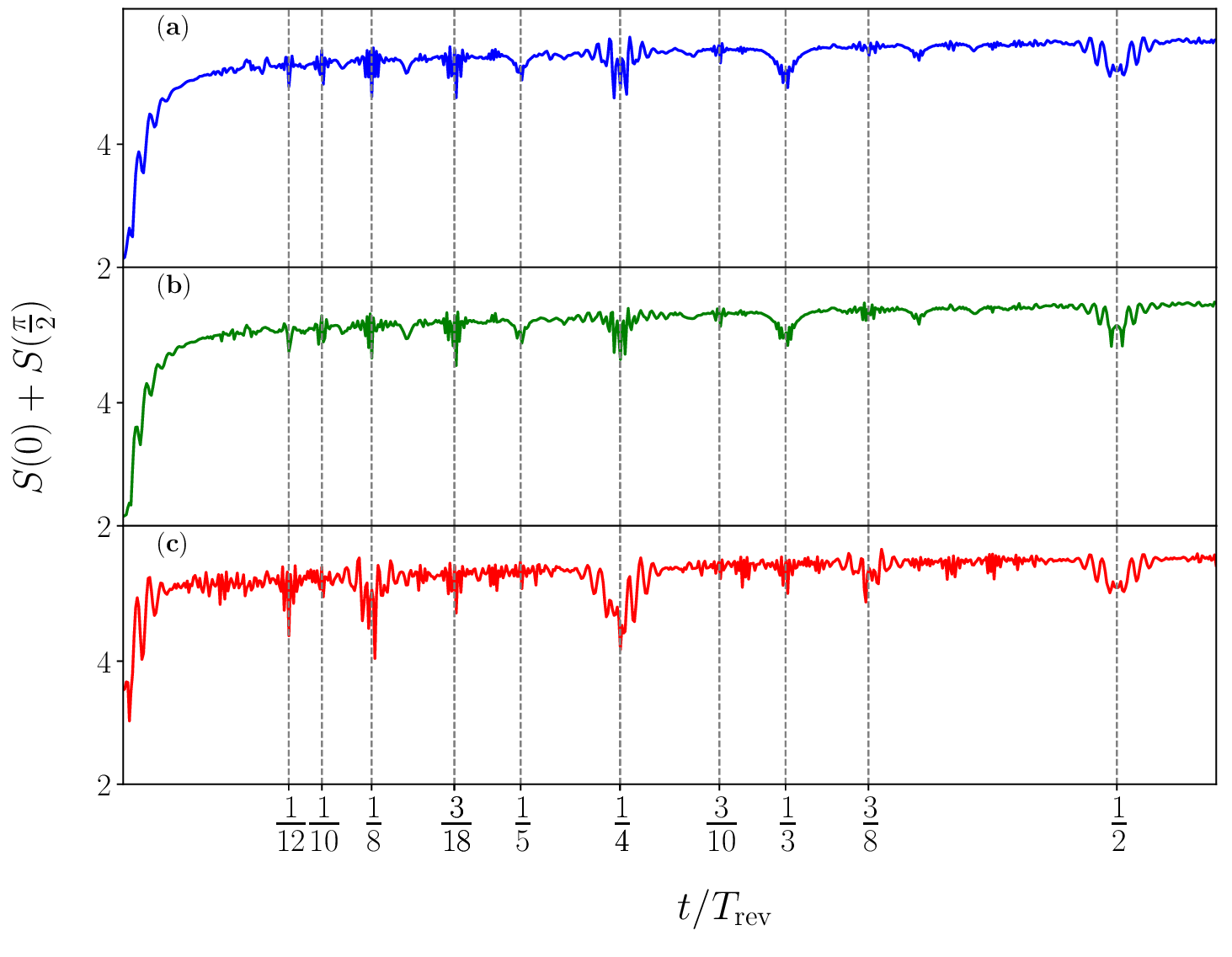}
    \caption{Dynamics of the sum of tomographic entropy in conjugate spaces for an initial (a) coherent state, (b) three-photon-added coherent state, and (c) even coherent state with $|\alpha|^2 = 40$ and $n_{\rm{max}}=100$, under amplitude damping with $\gamma = 0.05$. The relative minima of the curves correspond to the instants of near fractional revivals and phase-space rotations.}
    \label{fig8}
\end{figure*}
\begin{figure*}[h]
    \centering
    \includegraphics[width=0.8\textwidth]{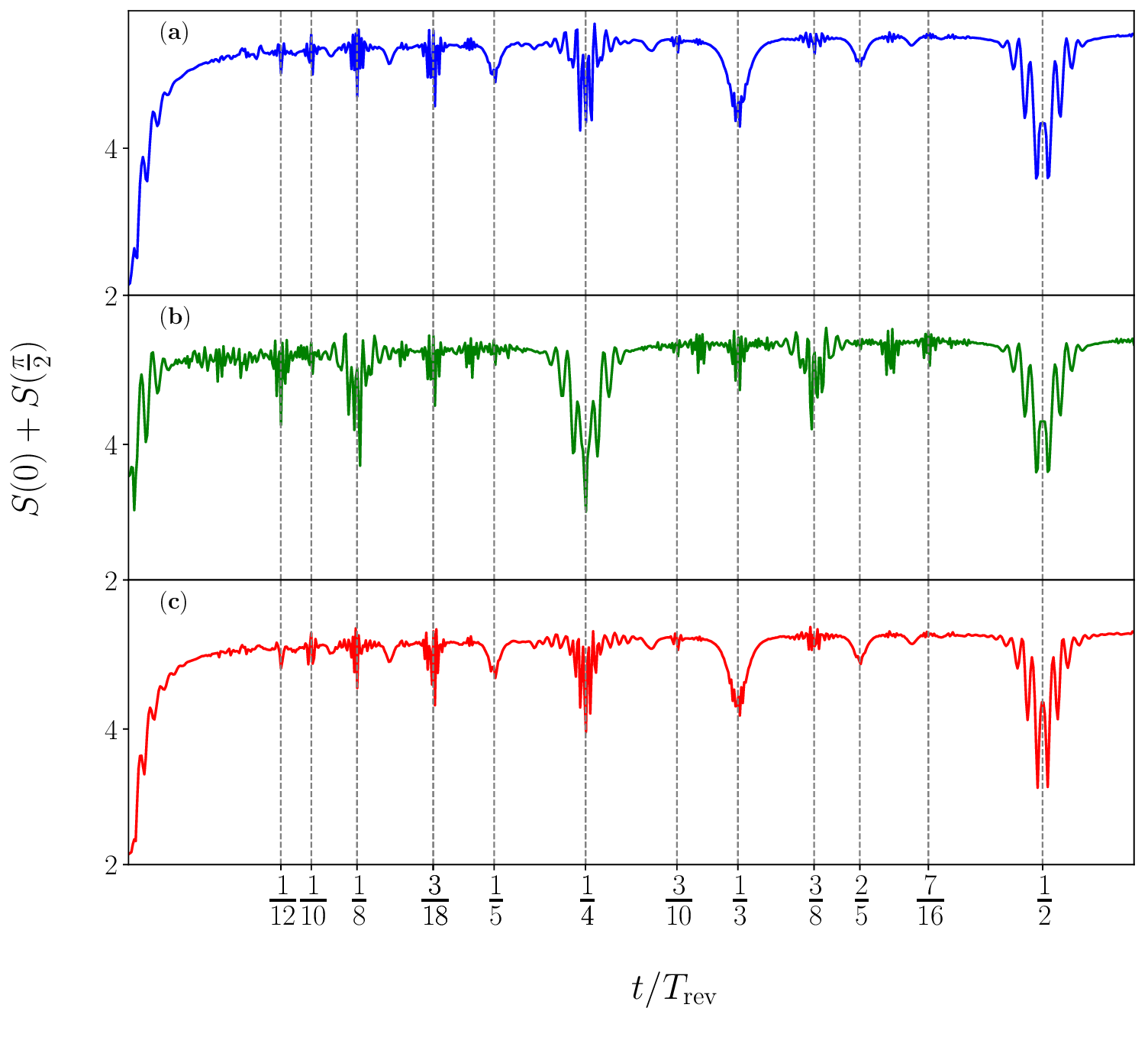}
    \caption{Dynamics of the sum of tomographic entropy in conjugate spaces for an initial (a) coherent state, (b) three-photon-added coherent state, and (c) even coherent state with $|\alpha|^2 = 40$ and $n_{\rm{max}}=100$, under phase damping with $\gamma = 0.05$. The relative minima of the curves correspond to the instants of near fractional revivals and phase-space rotations.}
    \label{fig9}
\end{figure*}

In the presence of environment-induced decoherence, the sum of tomographic entropies in conjugate spaces no longer returns to its initial value, regardless of the initial state. Fig.~\ref{fig8} illustrates the dynamics of tomographic entropies under weak amplitude damping with an interaction strength of $\gamma = 0.05$. The temporal evolution is shown for an initial coherent state, a three-photon-added coherent state, and an even coherent state with a field strength of $|\alpha|^2 = 40$, over a short interval ($t/T_{\mathrm{rev}} \leq 1/2$) to highlight the signatures of near fractional revivals. The relative minima in the entropy dynamics correspond to the instants of near fractional revivals, where the state is either a phase-rotated superposition of a finite number of initial wave packets or a phase-rotated version of the initial state itself. These signatures, indicated by the minima in the sum of tomographic entropies, demonstrate robustness against decoherence. However, as the interaction time increases, some fractional revivals are washed out from the dynamics and no longer produce clear signatures. For instance, the fractional revivals at $t = 2\trev/5$ and $7\trev/16$ observed in figure~\ref{fig7} are not distinctly visible in figure~\ref{fig8}. This highlights the gradual degradation of revival signatures with increasing decoherence effects. But for phase damping with weak coupling, the nonclassicality is largely preserved, as evident from figure~\ref{fig9}.
The sum entropy dynamics under phase damping resemble those of purely unitary evolution in the weak coupling regime for $0 \leq t \leq \trev/2$. For larger $\gamma$, the decay becomes significantly more pronounced. In the amplitude-damping model at zero temperature, photon loss drives the system to the vacuum state in the long-time limit $\gamma\,t \to \infty$. Consequently, the optical tomogram approaches that of the vacuum and the sum of tomographic entropies $S(0) + S(\pi/2)$ tends to $(1 + \ln\,\pi)$. For an initial coherent state, this may appear as though $S(0) + S(\pi/2)$ is returning to its initial value, since both the coherent state $\ket{\alpha}$ and the vacuum state $\ket{0}$ yield $S(0) + S(\pi/2) = (1 + \ln\,\pi)$. However, this does not correspond to revival and can be verified by plotting the optical tomogram of the state in the long-time limit. In contrast, phase damping preserves photon-number populations and only suppresses coherences, leading to a diagonal Fock-state mixture rather than the vacuum state. Accordingly, in this case, the entropy sum does not, in general, approach $(1+\ln\,\pi)$, and its asymptotic value depends on the initial state.

Overall, the relative minima in the dynamics of the sum of tomographic entropies in conjugate spaces provide a reliable indicator of revivals and higher-order fractional revivals, and these signatures remain robust against decoherence. Since tomographic entropy can be directly calculated from homodyne measurement data without requiring the reconstruction of the density matrix or the quasiprobability distribution, this theoretical approach may offer valuable insights for experimentalists performing high-precision measurements and identifying revivals and fractional revivals using balanced homodyne detection.

\section{Nonclassicality Dynamics in a Cubic medium}\label{Sec4}

The role of higher-order nonlinearities in nonclassicality has been an important arena of investigation, ranging from theoretical works \cite{Marek2011,Meyer2014} to experimental platforms \cite{Yukawa2013,Park2014}. In this section, we extend our investigation to the cubic nonlinear medium to examine the robustness of tomogram-based quantifiers in a higher-order nonlinearity. The analysis begins with the closed-system case ($\gamma = 0$), followed by a detailed study of open-system dynamics under both amplitude and phase damping channels. By evaluating the homodyne nonclassical area and tomographic entropy in these scenarios, we demonstrate how dissipation influences the manifestation of revivals, fractional revivals, and macroscopic superposition structures in a cubic nonlinear medium.
In a cubic Kerr medium, the system evolves under the nonlinear Hamiltonian in Eq.~(\ref{Hamilt_sys2}) with a cubic dependence on the photon number. The revival time under the cubic Hamiltonian $H_{\rm{cub}}$ is given by $\trev= \pi/\chi_2$.
If the initial state of the system is denoted by 
$\ket{\psi(0)}$, the time evolution under the unitary operator generated by the cubic Hamiltonian is described by
\begin{equation}
\ket{\psi(t)} = U(t) \ket{\psi(0)} = e^{-iH_{\rm{cub}}t} \ket{\psi(0)} 
\end{equation}
Expressing the initial state in the Fock basis, the evolved state under cubic nonlinearity becomes, 
\begin{equation}
\ket{\psi(t)}  = \sum_{n=0}^{\infty}C_n e^{-i\chi_2tn(n-1)(n-2)} \ket{n}. 
\end{equation}
Using Eq.~(\ref{fock_quad}), the optical tomogram of the state evolving in a cubic medium at any time t is expressed as, 
\begin{equation}
    \omega(X_{\theta},\theta,t)= \frac{e^{-X_{\theta}^2}}{\sqrt{\pi}} 
    \left| \frac{C_n e^{-i\chi_2 t n (n-1)(n-2)}}{\sqrt{n!} 2^{n/2}} e^{-in\theta} H_n(X_{\theta})\right|^2.
\end{equation}
Now we can proceed to investigate the dynamics of nonclassical area and tomographic entropy in a nonlinear cubic medium.

\subsection{Dynamics of nonclassical area}\label{Sec4.1}

The optical tomogram for unitary evolution in a cubic medium is given by, 
\begin{eqnarray}
\omega(X_\theta, \theta, t) & =\frac{e^{-X_{\theta}^2}}{\sqrt{\pi}}\sum_{n,m=0}^{\infty}
\frac{H_n(X_{\theta})H_m(X_{\theta})}{2^{(n+m)/2}\sqrt{n!m!}}\nonumber\\
&\times e^{-i(n-m)\theta}\,
\rho_{n, m}(t)
\label{optomo_cub}
\end{eqnarray}
where
\begin{equation}
\rho_{nm}(t)= e^{-i\chi(n(n-1)(n-2)-m(m-1)(m-2))t}\rho_{nm}(0).
\end{equation}

In contrast to a Kerr medium, the unitary evolution under the cubic Hamiltonian leads to fractional revivals of the initial state. In particular, three-component fractional revivals occur at \(\trev/3\) and \(2\trev/3\) within the interval 
\(0 \leq t \leq \trev\), while the complete reconstruction of the state occurs at 
\(t=\trev\).
Nonclassical area captures these instants by exactly imitating the nonclassical area of the initial states. The nonclassical area dynamics for a coherent state, a 3-photon added coherent state and an even coherent state evolving inside a cubic Kerr medium are analysed. 
\begin{figure}[h]
    \centering
    \includegraphics[width=0.5\textwidth]{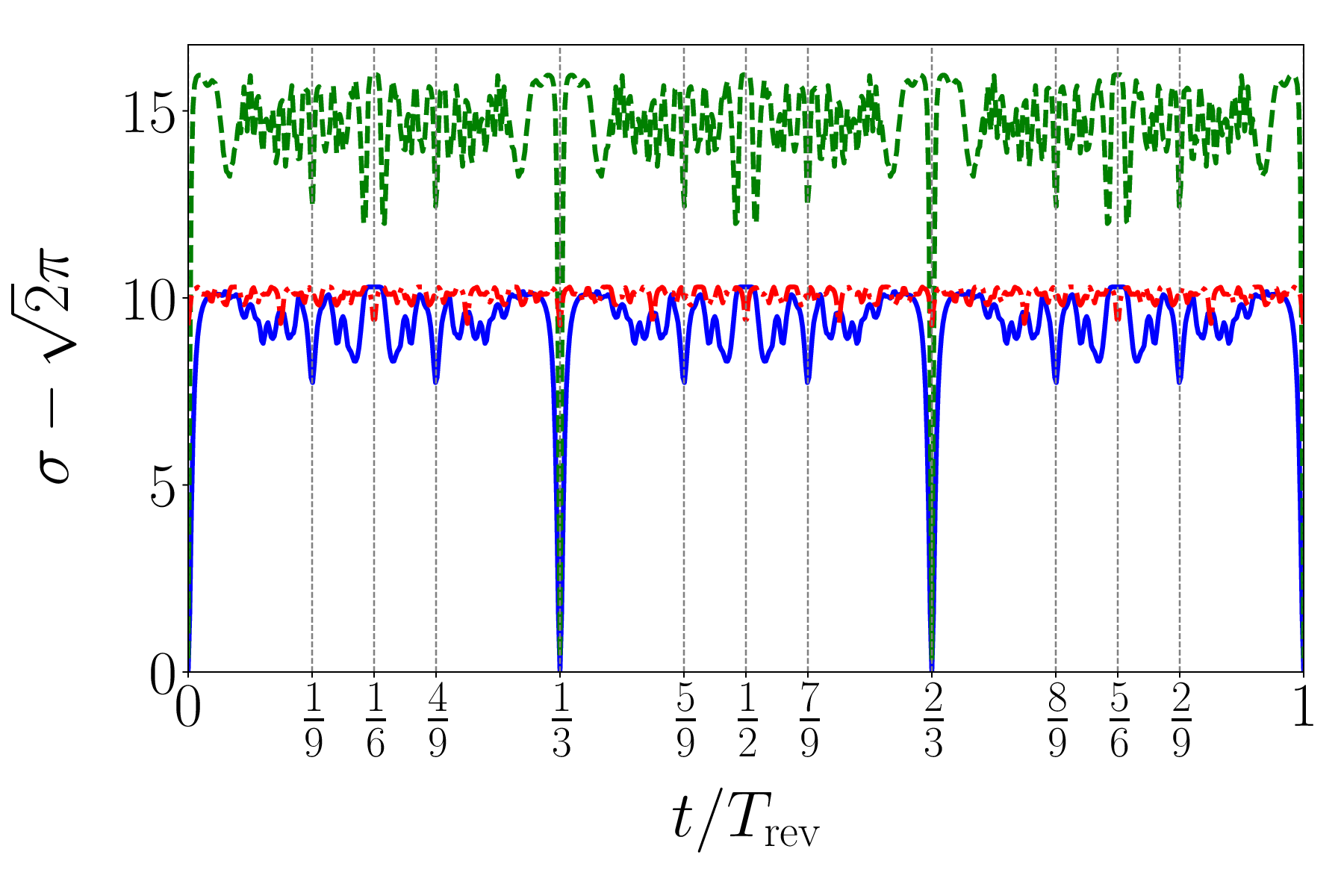}
    \caption{Dynamics of nonclassical area for an initial coherent state(solid),3-photon added coherent state(dashed)and even coherent state(dash-dot), with \(\modu{\alpha}^2\)= 5,  for \(\gamma=0\).}
    \label{fig10}
\end{figure}
\begin{figure}[h]
    \centering
    \includegraphics[width=0.5\textwidth]{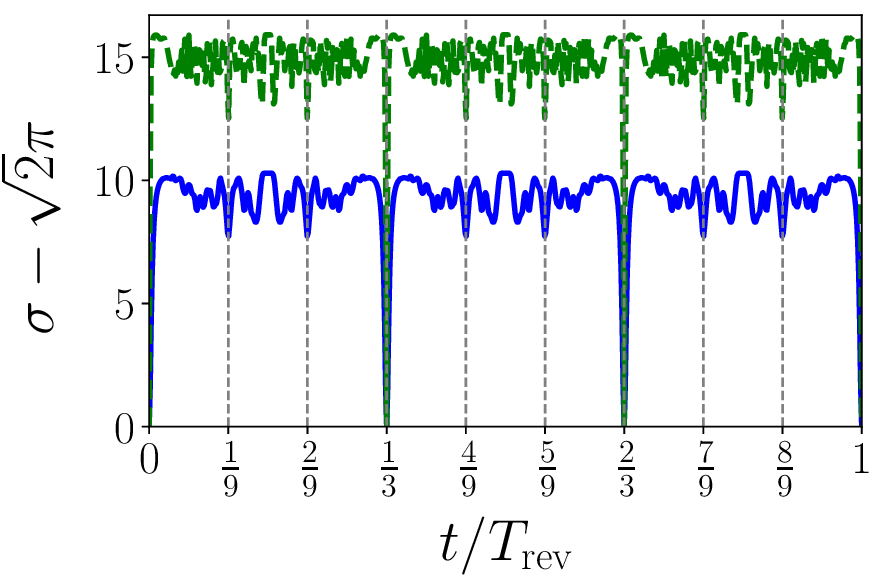}
    \caption{Dynamics of the nonclassical area for an initial coherent state with \(\modu{\alpha}^2\)= 5(solid), 10(dashed) for \(\gamma=0\).}
    \label{fig11}
\end{figure}
The nonclassical area computation is carried out for initial states with average photon number 
$\langle n\rangle=5$, in a Hilbert space dimension $n_{\rm{max}}=60$.
For \(\gamma=0\) the evolution is unitary with nonclassical area symmetric about \(t=\trev/2\) for  \(0 \leq t \leq\trev\). The nonclassical area returns to its initial value at the fractional-revival times \(t=\trev/3\)  and \(t=2\trev/3\), and again at the full revival time \(t=\trev\) within the interval \(0 \leq t \leq\trev\) such as coherent state has \(\sigma-\sqrt{2}\pi=0\) at the revivals (see figure~\ref{fig10}). Local minima are observed at instants like \(t=\trev/9\,, 4\trev/9\) for a coherent state, which indicate formation of macroscopic superposition states. The 3-photon added coherent state has a similar dynamics but has a higher nonclassical area owing to its initial inherent nonclassicality.
When the available number of photons in the initial state is higher, the greater will be its nonclassical area. For a coherent state, this can be observed from figure~\ref{fig11}. 
Now we can further investigate the nonclassical area dynamics in a cubic Kerr medium in distinct decoherence channels. 

\subsection{Amplitude Decay Model}\label{Sec4.2}

As said, amplitude damping induces both decay of coherence and population over time, while phase damping induces coherence decay only.
In the Kraus operator form, the state under amplitude damping in a cubic medium can be written as

\begin{equation}
\rho_t=\sum_{k=0}^{\infty}\frac{(1-e^{-\gamma t})^k}{k!}L_t U(t)a^k \rho(0){a^\dagger}^k U^\dagger (t)L_t^\dagger
\end{equation}
where $L_t=e^{-\gamma t a^\dagger a/2}$ and $U(t)=e^{-i\chi tn(n-1)(n-2)}$\\
The density matrix evolution equation is,
\begin{eqnarray}
\dot{\rho}_{n,m}(t) &=&\left(-i\chi (n(n-1)(n-2)-m(m-1)(m-2))\right.\nonumber\\
&&\left.-\frac{\gamma}{2} (n+m)\right)
\rho_{n,m}(t) +\gamma \sqrt{(n+1)(m+1)}\nonumber\\
&&\rho_{n+1,m+1}(t)
\end{eqnarray}
which on solving gives the elements as,
\begin{eqnarray}
\rho_{n,m}(t)&=& 
e^{-i\chi [n(n-1)(n-2) - m(m-1)(m-2)t]} \,
e^{-\gamma(n + m)t/2} \nonumber \\
&&\times \sum_{k=0}^{\infty} 
\binom{n + k}{k}^{1/2} \binom{m + k}{k}^{1/2}\nonumber \\ &&\frac{(1 - e^{-\gamma t})^k}{k!} 
\rho_{n + k, m + k}(0).
\end{eqnarray}
\begin{figure}[h]
\centering
\includegraphics[width=0.5\textwidth]{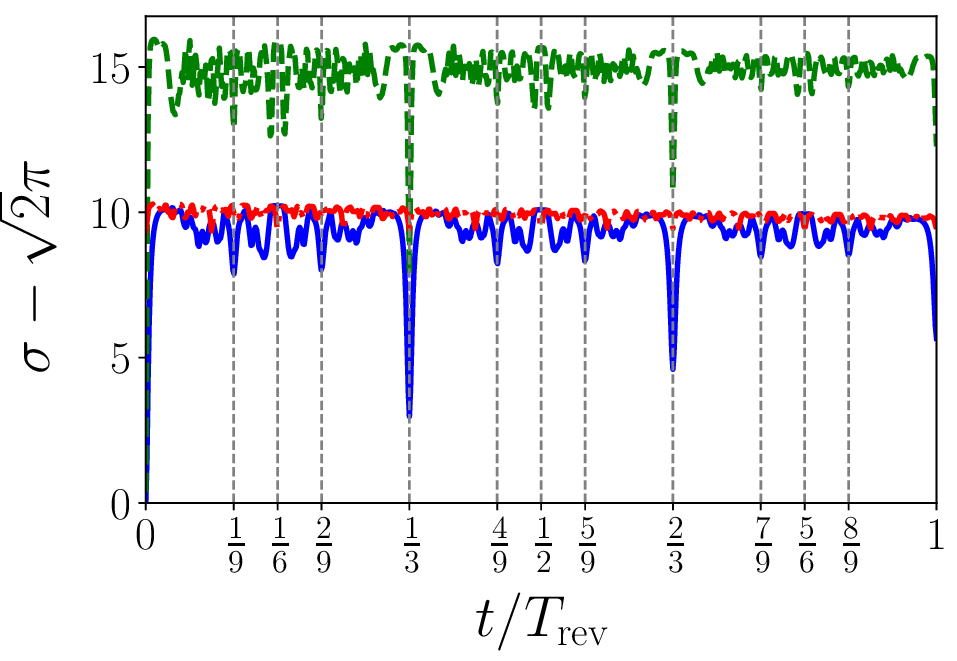}
\caption{Dynamics of nonclassical area under amplitude damping for a coherent state (dash-dot), three-photon-added coherent state (dash) and  even coherent state (solid line) with $|\alpha|^2 = 5$ with \(\gamma=0.1\). Hilbert space dimension $n_{\rm{max}}$ is set to $60$.}
\label{fig12}
\end{figure}
Hence the optical tomogram under amplitude damping in a cubic medium takes the form,
\begin{eqnarray}
\omega(X_\theta, \theta, t) & =\frac{e^{-X_{\theta}^2}}{\sqrt{\pi}}\sum_{n,m=0}^{\infty}
\frac{H_n(X_{\theta})H_m(X_{\theta})}{2^{(n+m)/2}\sqrt{n!m!}}\nonumber\\
& e^{-i(n-m)\theta}\,e^{-i\chi(n(n-1)(n-2)-m(m-1)(m-2))t}\nonumber\\
& e^{-\gamma(n+m)t/2}\sum_{k=0}^{\infty} 
\binom{n + k}{k}^{1/2} \binom{m + k}{k}^{1/2} \nonumber \\ 
& \frac{(1 - e^{-\gamma t})^k}{k!} 
\rho_{n + k, m + k}(0)
\label{optomo_cub_ampdcy}
\end{eqnarray}

\begin{figure*}[t]
    \centering
    \setlength{\tabcolsep}{2pt} 
    \renewcommand{\arraystretch}{0.5} 
    
    \begin{tabular}{ccccc}
    
        \rotatebox{90}{$\theta $}
        \begin{minipage}{0.22\textwidth}
            \centering
            \textbf{(i)} $\gamma t=0.01$\\
            \includegraphics[width=\linewidth]{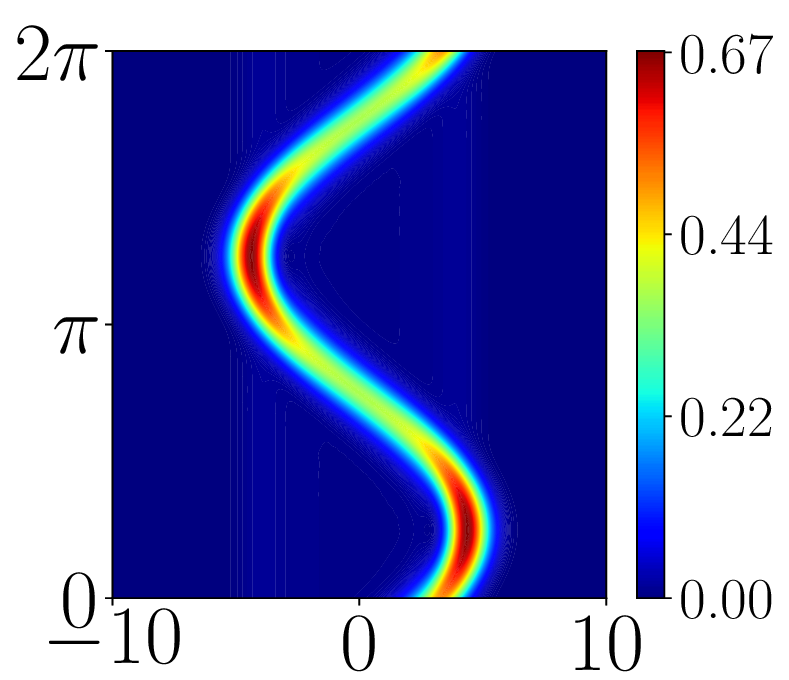}\\
            $X_{\theta}$
        \end{minipage}
        &
        \begin{minipage}{0.22\textwidth}
            \centering
            \textbf{(ii)}$\gamma t=0.1$\\
            \includegraphics[width=\linewidth]{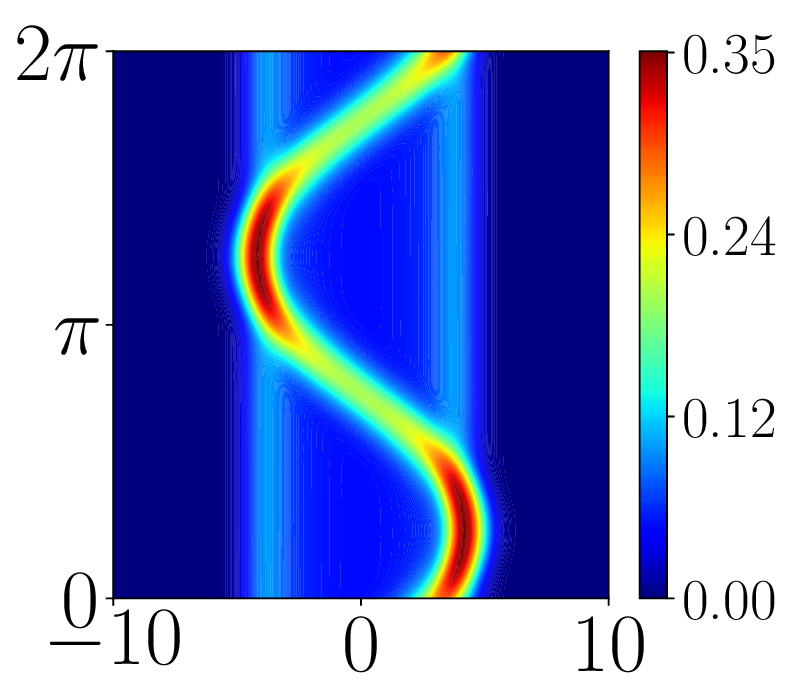}\\
            $X_{\theta}$
        \end{minipage}
        &
        \begin{minipage}{0.22\textwidth}
            \centering
            \textbf{(iii)}$\gamma t=1$\\
            \includegraphics[width=\linewidth]{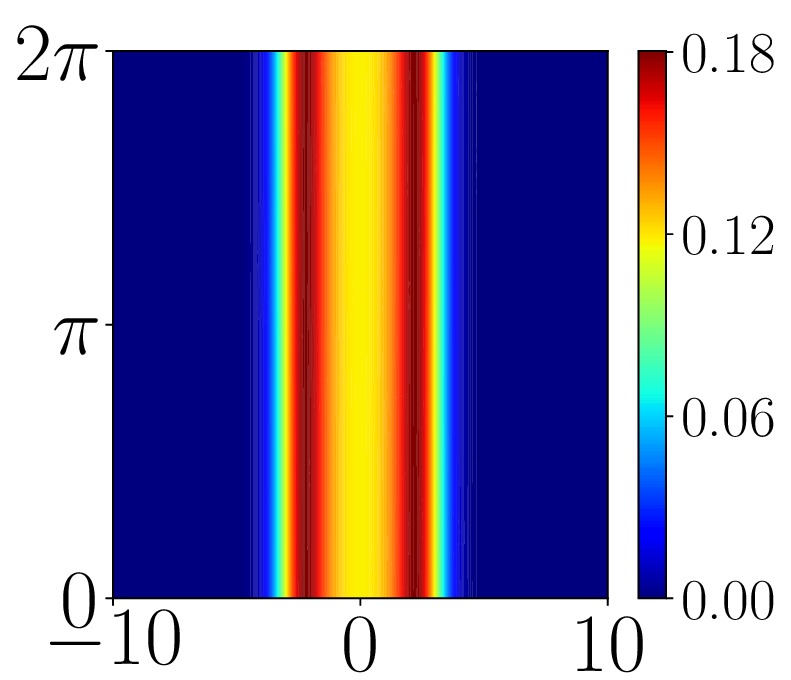}\\
            $X_{\theta}$
        \end{minipage}
        &
        \begin{minipage}{0.22\textwidth}
            \centering
            \textbf{(iv)}$\gamma t \to \infty$\\
            \includegraphics[width=\linewidth]{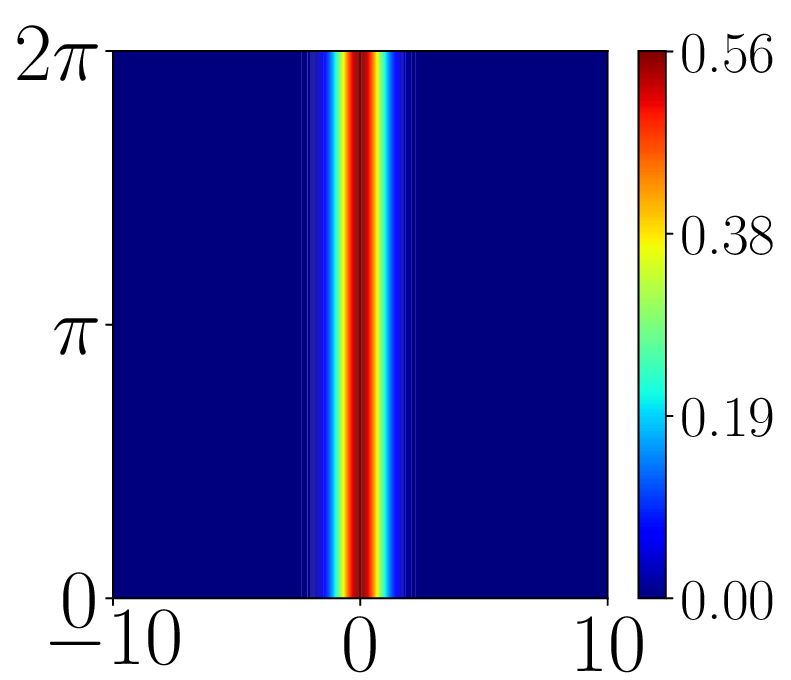}\\
            $X_{\theta}$
        \end{minipage}
    \end{tabular}
    \caption{Optical tomogram at revival in the presence of amplitude damping in a cubic nonlinear medium for a 3-photon added coherent state with \(\modu{\alpha}^2\)= 5. Hilbert space dimension is kept at $n_{max}=60$}
    \label{fig13}
\end{figure*}
Amplitude damping causes gradual suppression of nonclassical area oscillations, indicating decay. The behaviour of the nonclassical area with amplitude damping for different initial states is plotted in figure~\ref{fig12}. No exact revival patterns are observed in the nonclassical area dynamics. Local minimas in the nonclassical area are signatures of near revivals and fractional revivals occurring under weak coupling. The photon added coherent state maintains a higher nonclassical area than the coherent and even coherent state for most of the time evolution, indicating resilience to damping. 

\begin{figure}[h]
\centering
\includegraphics[width=0.5\textwidth]{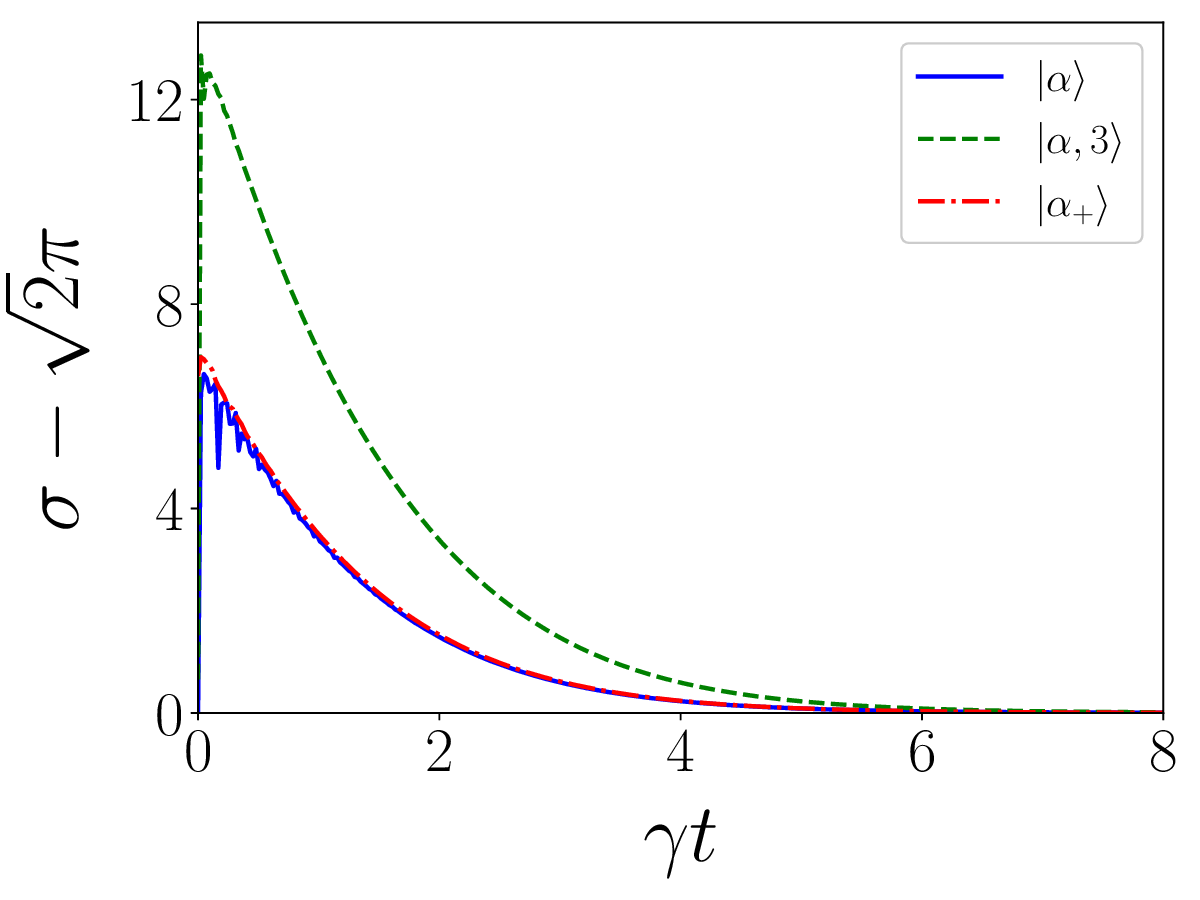}
\caption{Evolution of the nonclassical area as a function of \(\gamma t\) with amplitude damping shown for an initial coherent state (solid line), a 3-photon-added coherent state (dashed line), and an even coherent state (dash-dot line) with \(\modu{\alpha}^2=3\)}
\label{fig14}
\end{figure}
The exponential death of nonclassicality towards that of a vacuum state under amplitude decay in the nonlinear cubic medium is manifested in the dynamics of the nonclassical area as shown in figure~\ref{fig14}. Amplitude damping exponentially decreases the nonclassical area, driving the system to a steady state characterized by zero nonclassical area. Thus, a vacuum steady state is achieved. Amplitude damping can be visualized in the corresponding optical tomograms of the respective states. Figure~\ref{fig13} presents the optical tomogram evolution of a 3-photon added coherent state in an amplitude decay channel. 
\subsection{Phase Decay Model}\label{Sec4.3}

Under phase decay in a cubic medium, the state is obtained as,
\begin{eqnarray}
\rho_{nm}(t)&=& e^{-i\chi(n(n-1)(n-2)-m(m-1)(m-2))t}
\nonumber\\
&&e^{-\gamma(n-m)^2 t/2}\rho_{nm}(0).
\end{eqnarray}
Hence the optical tomogram becomes,
\begin{eqnarray}
\omega(X_\theta, \theta, t) & =\frac{e^{-X_{\theta}^2}}{\sqrt{\pi}}\sum_{n,m=0}^{\infty}
\frac{H_n(X_{\theta})H_m(X_{\theta})}{2^{(n+m)/2}\sqrt{n!m!}}\nonumber\\
& e^{-i(n-m)\theta}\,e^{-i\chi(n(n-1)(n-2)-m(m-1)(m-2))t}\nonumber\\
& e^{-\gamma(n-m)^2 t/2}\rho_{nm}(0).
\end{eqnarray}
\begin{figure}[h]
\centering
\includegraphics[width=0.45\textwidth]{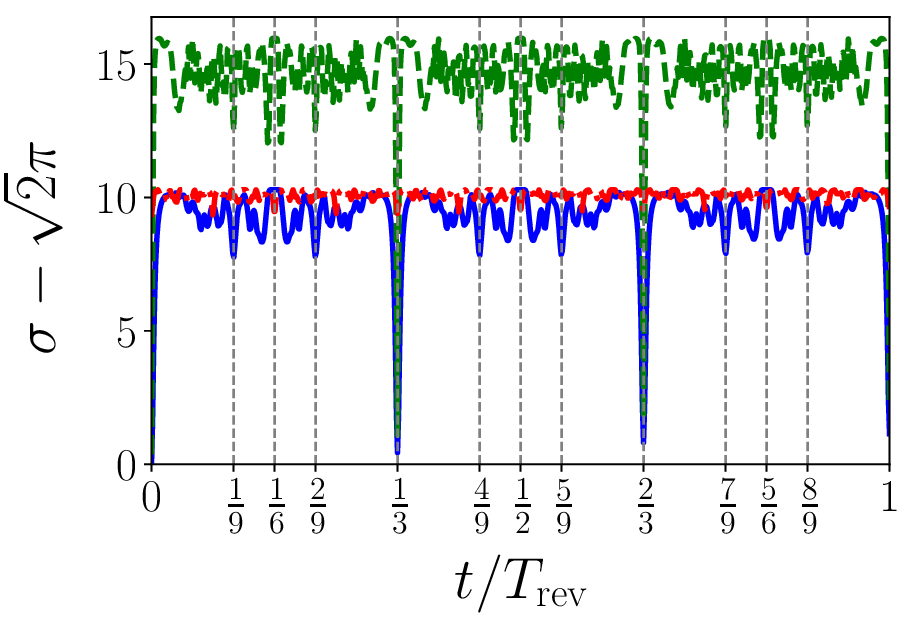}
\caption{Time evolution of the nonclassical area with phase damping for \(\gamma = 0.1\) and \(\modu{\alpha}^2=5\) , shown for an initial coherent state (solid line), a 3-photon-added coherent state (dashed line), and an even coherent state (dash-dot line). Hilbert space dimension $n_{\rm{max}}$ is set to $60$.}
\label{fig15}
\end{figure}

\begin{figure*}[h]
    \centering
    \setlength{\tabcolsep}{2pt} 
    \renewcommand{\arraystretch}{0.5} 
    
    \begin{tabular}{ccccc}
    
        \rotatebox{90}{${\theta}$}
        \begin{minipage}{0.22\textwidth}
            \centering
            \textbf{(i)} $\gamma t=0.01$\\
            \includegraphics[width=\linewidth]{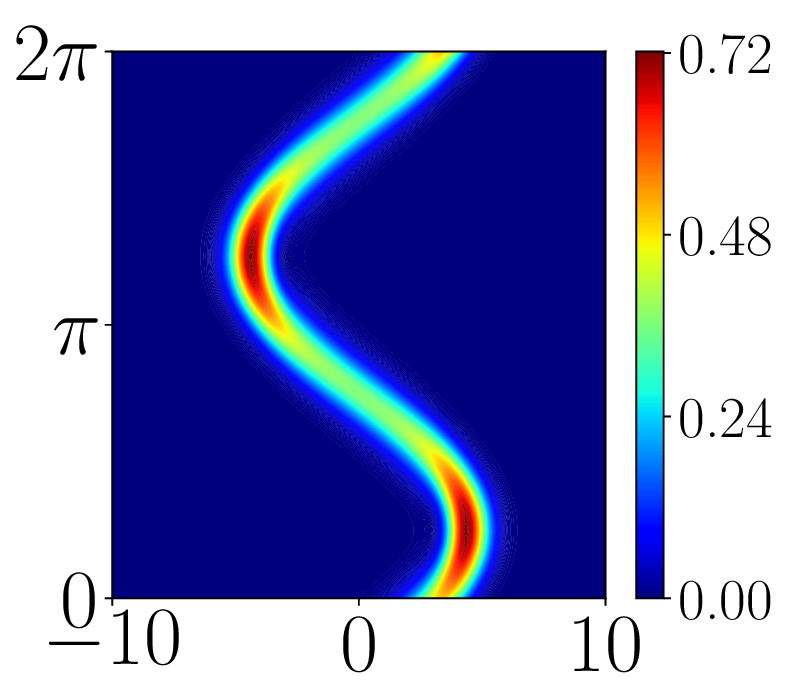}\\
            $X_{\theta}$
        \end{minipage}
        &
        \begin{minipage}{0.22\textwidth}
            \centering
            \textbf{(ii)}$\gamma t=0.1$\\
            \includegraphics[width=\linewidth]{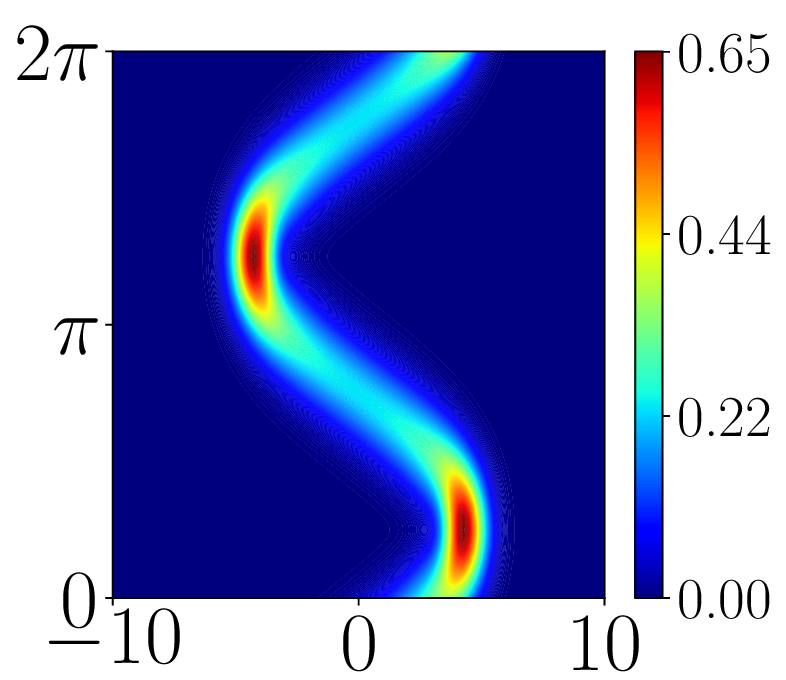}\\
            $X_{\theta}$
        \end{minipage}
        &
        \begin{minipage}{0.22\textwidth}
            \centering
            \textbf{(iii)}$\gamma t=1$\\
            \includegraphics[width=\linewidth]{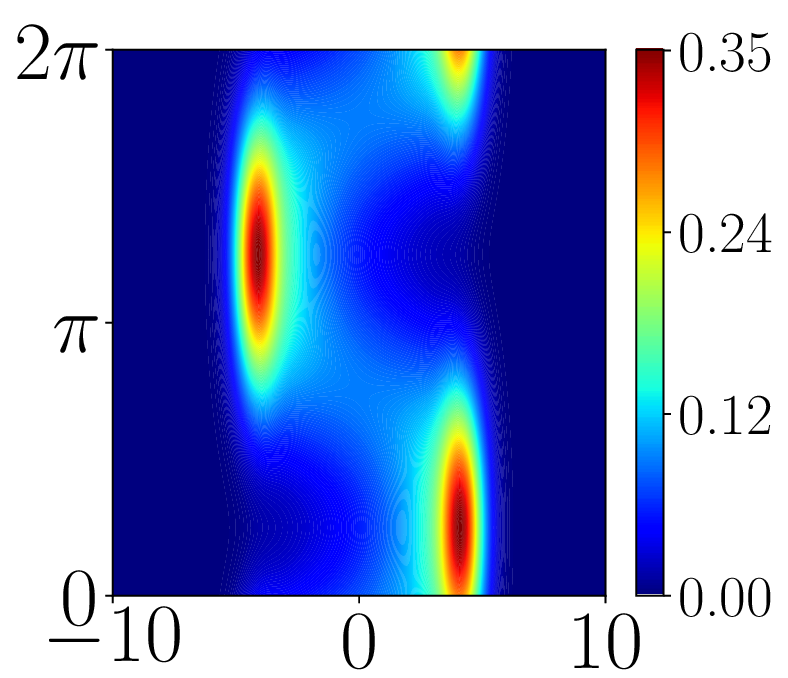}\\
            $X_{\theta}$
        \end{minipage}
        &
        \begin{minipage}{0.22\textwidth}
            \centering
            \textbf{(iv)}$\gamma t \to \infty$\\
            \includegraphics[width=\linewidth]{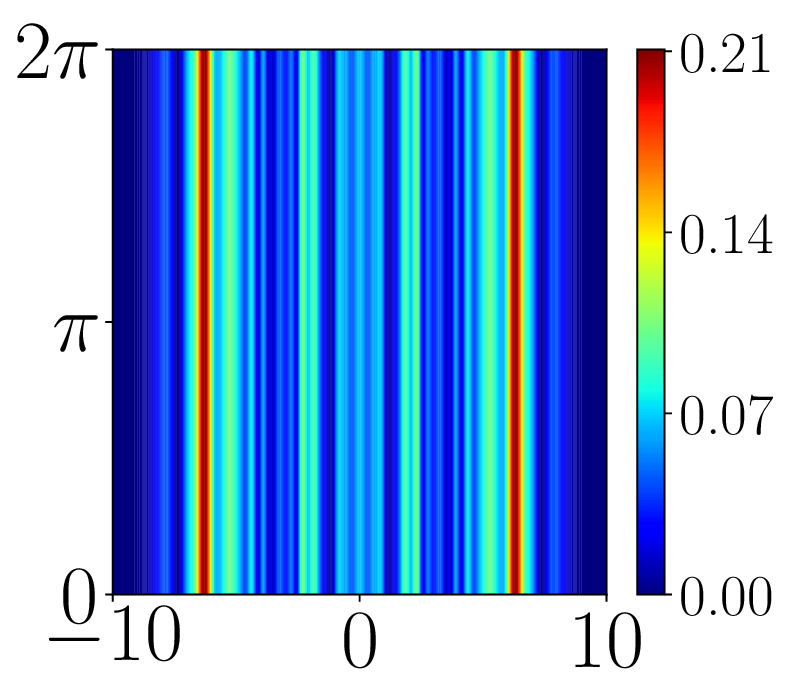}\\
            $X_{\theta}$
        \end{minipage}
    \end{tabular}
    \caption{Optical tomogram at revival in the presence of phase damping in a cubic nonlinear medium for a 3-photon added coherent state with \(\modu{\alpha}^2\)= 5. Hilbert space dimension is kept at $n_{max}=60$}
    \label{fig16}
\end{figure*}
\begin{figure}[h]
\centering
\includegraphics[width=0.5\textwidth]{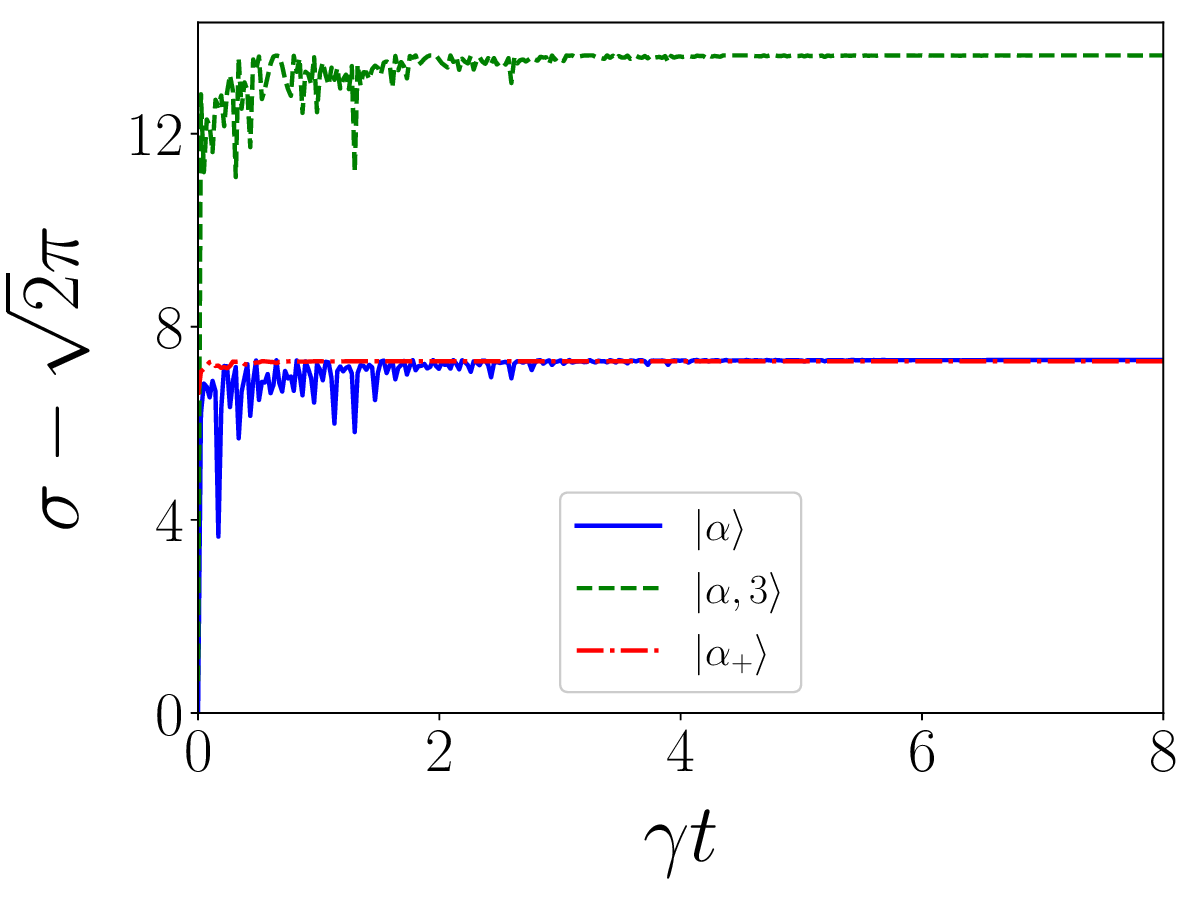}
\caption{Evolution of the nonclassical area as a function of \(\gamma t\) with phase damping shown for an initial coherent state (solid line), a 3-photon-added coherent state (dashed line), and an even coherent state (dash-dot line) with \(\modu{\alpha}^2=5\)}
\label{fig17}
\end{figure}

Phase decay is less destructive to the nonclassicality of the state compared to amplitude decay. Though weakened, the minimas are still visible in the nonclassical area for pure dephasing. This can be observed in figure~\ref{fig15}, which plots the nonclassical area dynamics of the phase damping model with coupling strength \(\gamma=0.1\). For the time range \(0\leq t\leq \trev\), the minimas are only very minimally destroyed, indicating that in the weakly coupled regime, near revivals and fractional revivals are still observed as local minimas in the nonclassical area. But as the interaction time is increased, the nonclassical features are gradually suppressed. Optical tomograms themselves capture the pure dephasing.  Figure~\ref{fig16} shows the evolution of the optical tomogram for a 3-photon added coherent state under phase damping. The background noise reflects the onset of decoherence, which ultimately drives the system to its steady state. The tomogram-derived nonclassical area captures the phase decay dynamics of the system (see figure~\ref{fig17}). All the coherence in the system is lost, while the population is preserved [$\rho_{n,n}(t)= \rho_{n,n}(0)$] under phase damping, eventually leading to the saturation of the nonclassical area.
It is important to note that this nonzero value does not imply nonclassicality, but rather reflects contributions from both classical and quantum uncertainties.
\subsection{Dynamics of tomographic entropy}\label{Sec4.4}

For an initial coherent state \(\ket{\alpha}\) propagating through a cubic medium without decoherence, the tomogram at instant t is
\begin{eqnarray}
\omega_\alpha(X_\theta, \theta, t) &=& \frac{e^{-|\alpha|^2} \, e^{-X_\theta^2}}{\sqrt{\pi}}\times \sum_{n=0}^{\infty} \frac{\alpha^n}{n! \, 2^{n/2}} \nonumber \\
&& \left|  e^{-i \chi t n(n-1)(n-2)}
\, e^{-i n \theta} \, H_n(X_\theta) \right|^2
\end{eqnarray}
For an initial p-photon added coherent state 
\(\ket{\alpha,p}\) in a cubic medium,

\begin{eqnarray}
\omega_{\alpha,p}(X_\theta, \theta, t) &=& 
\frac{e^{-|\alpha|^2}}{p! \, L_p(-|\alpha|^2)} \cdot \frac{e^{-X_\theta^2}}{\sqrt{\pi}} \nonumber \\
&& \bigg|\sum_{n=p}^{\infty} 
\frac{\alpha^{n-p} \, e^{-i \chi t n(n-1)(n-2)}}{(n-p)! \, 2^{n/2}} 
\nonumber\\
&& \times e^{-i n \theta} \, H_n(X_\theta)\bigg|^2
\end{eqnarray}

An even coherent state \(\ket{\alpha_+}\) evolving inside a cubic medium has optical tomogram,

\begin{eqnarray}
\omega_{+}(X_\theta, \theta, t) &&= 
4 N_{+}^2 \, \frac{e^{-|\alpha|^2} \, e^{-X_\theta^2}}{\sqrt{\pi}} \times \bigg| \sum_{n=0}^{\infty} 
\frac{\alpha^n}{n! \, 2^{n/2}}  \nonumber \\
&&
e^{-i \chi t n(n-1)(n-2)} e^{-i n \theta} \, H_n(X_\theta)  \delta_{\left[\frac{n}{2}\right], \frac{n}{2}} 
\bigg|^2.
\end{eqnarray}

\begin{figure*}[h]
    \centering
    \includegraphics[width=0.8\textwidth]{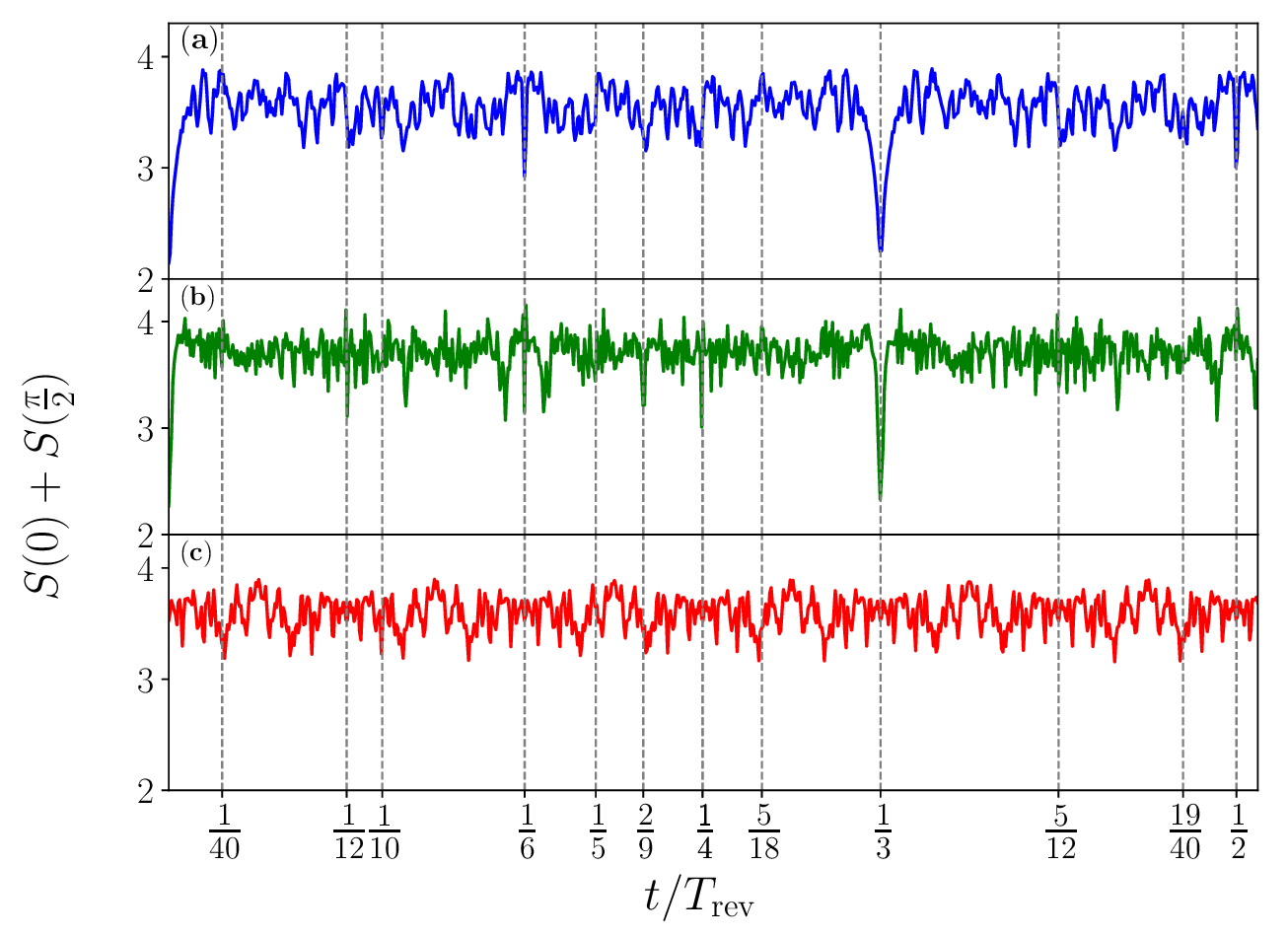}    
    \caption{Dynamics of the sum tomographic entropies of an initial (a) coherent state, (b) three-photon-added coherent state, and (c) even coherent state with $|\alpha|^2=5$, in the absence of decoherence ($\gamma=0$). Relative minima in the curves correspond to the instants of fractional revivals and phase-space rotations. Hilbert space dimension is set to $70$.}
    \label{fig18}
\end{figure*}
\begin{figure*}[h]
    \centering
    \includegraphics[width=0.8\textwidth]{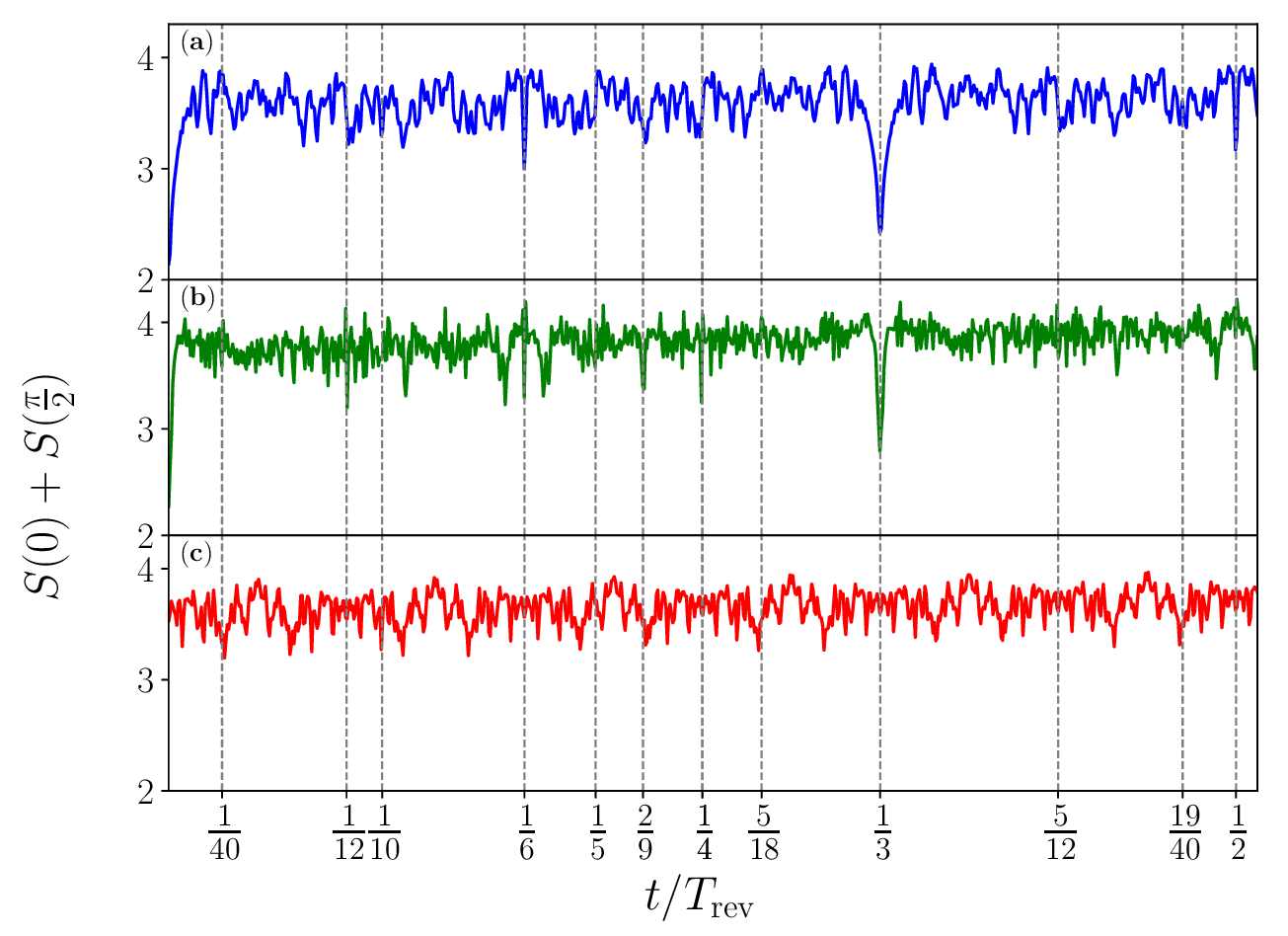}    
    \caption{Dynamics of the sum tomographic entropies of an initial (a) coherent state, (b) three-photon-added coherent state, and (c) even coherent state with $|\alpha|^2=5$, under amplitude damping ($\gamma=0.05$). Relative minima in the curves correspond to the instants of fractional revivals and phase-space rotations. Hilbert space dimension is set to $70$.}
    \label{fig19}
\end{figure*}
\begin{figure*}[h]
    \centering
    \includegraphics[width=0.8\textwidth]{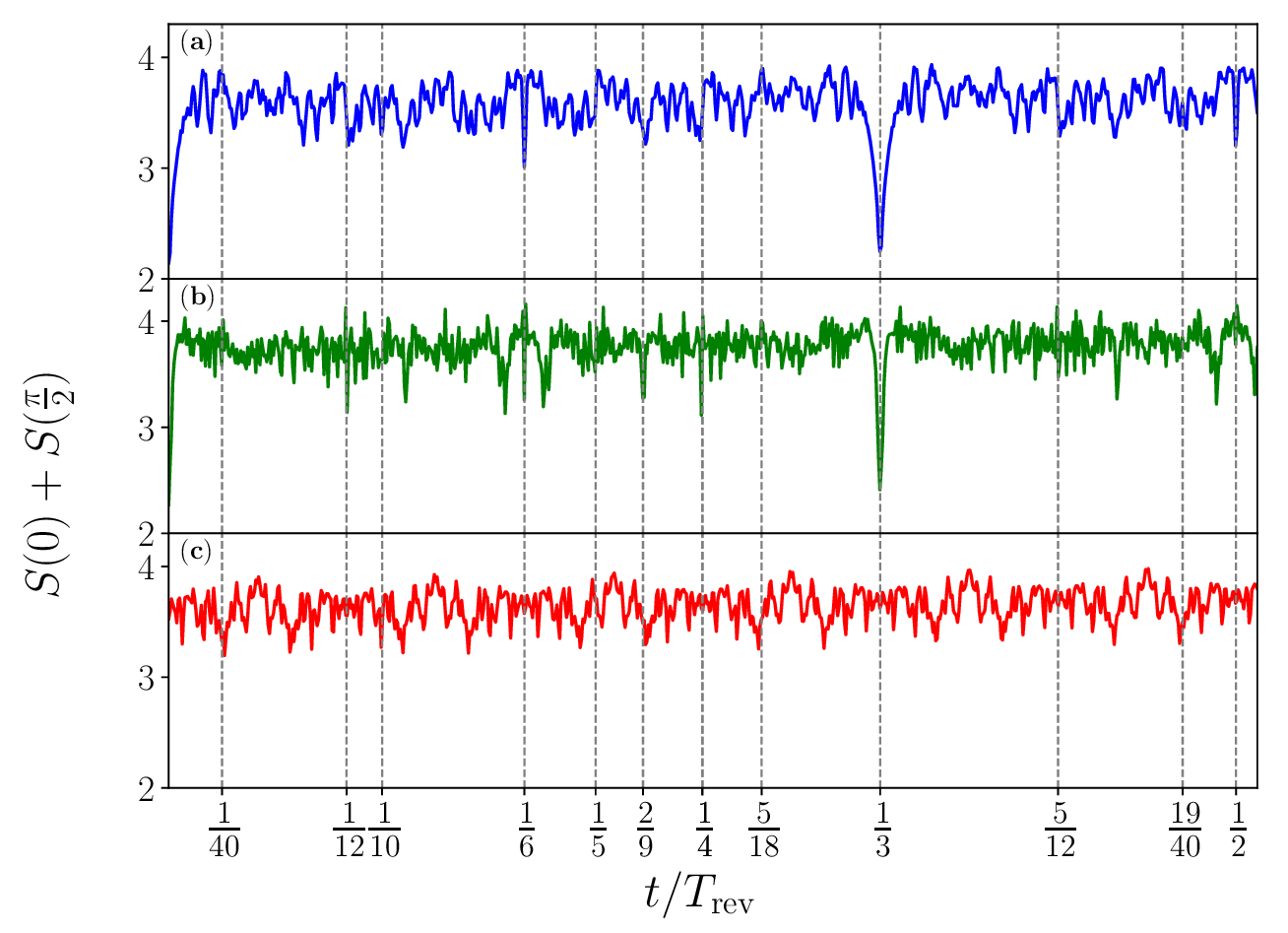}    
    \caption{Dynamics of the sum tomographic entropy of an initial (a) coherent state, (b) three-photon-added coherent state, and (c) even coherent state with $|\alpha|^2=5$, under phase damping ($\gamma=0.05$). Relative minima in the curves correspond to the instants of fractional revivals and phase-space rotations. Hilbert space dimension is set to $70$.}
    \label{fig20}
\end{figure*}
The higher-order fractional revivals that are not captured in the nonclassical area are captured in the sum tomographic entropy. The tomographic entropy obtained as per Eq.~(\ref{tomogramentropy}) obeys the uncertainty relation in Eq.~(\ref{entropy_uncertainty}). For the cubic nonlinear medium with mean photon number $|\alpha|^2 = 5$, convergence checks were carried out for coherent, even coherent, and 3-photon added coherent states, and a cutoff $n_{\rm{\max}} = 60$ was found to yield stable results. In all cases, relevant observables were verified to remain invariant under further increase of the cutoff, confirming that the chosen truncations provide numerically converged results and that the fine structures in the entropy are not artifacts of the finite Hilbert space. The time evolution was computed over the interval \(t\in [0, 0.55 \trev]\). A uniform grid of 700 time points was used to adequately resolve higher-order fractional revival structures.

The optical tomogram was evaluated over the quadrature range \(X_{\theta} \in [-10,10] \), discretized using 200 grid points. The sum tomographic entropy dynamics across conjugate spaces for an entirely isolated system for three different initial states in a cubic medium is shown in figure~\ref{fig18}. When the evolution is unitary, a state emerging from any of the initial state has its sum tomographic entropy returning to its initial value at each complete revival. The sum tomographic entropy for all three initial states (coherent, three-photon-added coherent, and even coherent) exhibits clear periodic minima. These minima correspond to the instants of fractional and full revivals in the quantum state, indicating the reformation of coherent or structured superposition states. The presence of local minima at fractional revival times signals the formation of macroscopic superpositions (Schr\"{o}dinger cat-like states), which are more nonclassical and thus have lower entropy. 

When coupling is introduced via amplitude damping, the periodicity and depth of the entropy minima are significantly reduced. The sharp features associated with revivals and fractional revivals are washed out, reflecting the loss of coherence and the transition toward classicality (See figure~\ref{fig19}). We have plotted the sum tomographic entropy under damping for \(\gamma=0.05\). As damping increases, the local minima become less pronounced, showing that the system's ability to form macroscopic superpositions is suppressed.  In the long time limit \(\gamma t \to \infty\), the sum of tomographic entropies
\( S(0) + S(\pi/2)\) approaches the value \(1 + \ln \pi\) under amplitude damping, corresponding to the vacuum state \(\ket{0}\).

Under phase damping (\(\gamma=0.05\)), the sum tomographic entropy still displays signatures of revivals and fractional revivals, though the features are less sharp than in the unitary case. The system retains some memory of its initial coherence, as indicated by the persistence of local minima. This suggests that phase damping, while degrading quantum coherence, does not destroy the revival structure in the weakly coupled regime. The dynamics of sum tomographic entropy for the three initial states in a cubic medium under phase damping is shown in figure~\ref{fig20}. The entropy curves for all three states under phase damping lie between the unitary and amplitude damping cases, highlighting that dephasing is less destructive to revival phenomena than amplitude damping.

\section{Conclusion}\label{Sec5}
In this work, we have established tomogram-based quantifiers, homodyne nonclassical area, and the sum tomographic entropy as powerful tools for identifying quantum signatures during the evolution of states. Unlike traditional measures such as Wigner negativity or Mandel's Q-parameter, which are often experimentally intractable or reconstruction-heavy, these tomogram-based measures bypass full density matrix reconstruction, thereby reducing numerical and experimental errors and making them readily applicable in experimental scenarios via balanced homodyne detection. In photonic systems, nonlinear models are ideal platforms to generate nonclassicality. We investigated the evolution of three representative states - a coherent state, a 3-photon added coherent state, and an even coherent state in both Kerr and cubic nonlinear media. We analysed nonclassicality dynamics in open quantum systems by incorporating decoherence through Lindblad master equations at zero temperature, considering both amplitude damping and phase damping channels. 

The nonclassical area, which quantifies the area projected by the optical tomogram relative to a coherent state, clearly reveals the occurrence of full and fractional revivals during unitary evolution in both the Kerr and cubic nonlinear media. Revivals are indicated by complete retrievals of the nonclassical area to its initial value, while fractional revivals are observed as pronounced dips, especially around the two-subpacket superposition time for isolated systems. Amplitude damping drives the system to vacuum, which manifests as an exponential decay in the nonclassical area for both Kerr and cubic nonlinear media. In contrast, phase damping gradually erodes quantum coherence while preserving the population distribution, ultimately leading to the saturation of the nonclassical area at long times. The nonclassical area is not a strict witness for nonclassicality owing to quantum-classical mixing. For the same amount of coupling, nonclassical area dynamics reveals that amplitude damping destroys coherence more rapidly than phase damping. These contrasting behaviours between damping types demonstrate the sensitivity of the nonclassical area to different decoherence mechanisms. To illustrate this effect, the optical tomograms of a 3-photon added coherent state subjected to amplitude and phase damping are presented.

In addition, we employed sum tomographic entropy, derived from the optical tomogram across conjugate quadratures, to investigate the formation of higher-order macroscopic superposition states. The sum tomographic entropy revealed signatures of revivals and higher-order fractional revivals through minimas and local minimas under unitary evolution. Complementing the behaviour of the nonclassical area, the sum tomographic entropy also provides clear signatures of amplitude and phase damping, evident through the gradual suppression of its characteristic local minima. In the weak coupling regime (\(\gamma \ll 1\)), the fidelity of the tomographic approach remains strong in tracking fine quantum structures. Our results demonstrate that the homodyne nonclassical area and sum tomographic entropy provide an experimentally accessible, real-time, and physically intuitive framework for characterizing the resilience and decay of nonclassical features in nonlinear open quantum systems, offering a compelling and practical alternative to conventional phase-space measures.





\bmhead{Acknowledgements}
M. R. thank the financial support provided for this research by the DST Anusandhan National Research Foundation (ANRF), Government of India, through the State University Research Excellence scheme, with reference number SUR/2022/003354. K. M. A. acknowledges the financial support from the Department of Science and Technology (DST), Government of India, through the INSPIRE Fellowship with reference number DST/INSPIRE/03/2025/000257 [IF240166].
\section*{Declarations}

\begin{itemize}

\item Competing interests 
\\The authors declare no conflict of interest.
\item Data availability 
\\ This manuscript has no associated data. No data is used to produce any result in the paper. All the figures in the manuscript are produced using the equations derived in the manuscript.
\end{itemize}

\noindent
\bibliography{sn-bibliography}

\end{document}